\documentclass[a4paper,12pt]{article}

\usepackage{amsmath}
\usepackage{amssymb}
\usepackage{amsthm}
\usepackage{mathrsfs}
\usepackage[T1]{fontenc}
\usepackage{enumerate}
\usepackage{color}

\def\mtr{matri}
\def\eqn{equation}

\def\tfn{transformation}

\def\fn{function}

\def\pl{Poisson--Lie}
\def\JL{Jacobi--Lie}
\def\jltpy{Jacobi--Lie T-plurality}

\def\dd{Drinfel'd double}

\def\4diml{four-dimensional}
\def\bkg{background}

\def\crspto{corresponding to\ }

\def\-1{^{-1}}
\def\half{\frac{1}{2}}
\def\coor{coordinate}
\def\real{\mathbb{R}}

\def\cd{{\mathfrak d}}
\def\cg{{\mathfrak g}}
\def\tcg{\tilde{\mathfrak g}}

\def\wh{\widehat}

\def\PL{Poisson--Lie }
\def\pltp{Poisson--Lie T-pluralit}
\def\pltd{Poisson--Lie T-dualit}

\def\JLtp{Jacobi--Lie T-pluralit}

\def\gsugra{Generalized Supergravity Equation}
\def\usugra{usual Supergravity Equation}
\def\sugra{Supergravity Equation}

\def\cf{{\mathcal {F}}}

\newcommand{\DDp}{DD$^+$}
\newcommand{\Exp}[1]{\operatorname{e}^{#1}}

\newcommand{\abs}[1]{\lvert {#1} \rvert}

\newcommand{\unit}{\mathbf{1}}
\newcommand{\nul}{\mathbf{0}}

\newcommand{\D}{\mathscr{D}}
\newcommand{\G}{\mathscr{G}}
\newcommand{\tG}{\widetilde{\mathscr{G}}}

\newcommand{\J}{\mathcal{J}}

\newcommand{\cB}{\mathcal B}
\newcommand{\cD}{\mathcal D}
\newcommand{\cE}{\mathcal E}\newcommand{\cF}{\mathcal F}
\newcommand{\cG}{\mathcal G}\newcommand{\cH}{\mathcal H}

\newcommand{\cW}{\mathcal W}

\begin{document}
\title{Plane-parallel waves as Jacobi-Lie models}

\author{Ivo Petr\footnote{ivo.petr@fit.cvut.cz}
\\ {\em Faculty of Information Technology,}
\\ {\em Czech Technical University in Prague,}
\\ {\em Czech Republic}
\and
Ladislav Hlavat\'y\footnote{ladislav.hlavaty@fjfi.cvut.cz}
\\ {\em Faculty of Nuclear Sciences and Physical Engineering,}
\\ {\em Czech Technical University in Prague,}
\\ {\em Czech Republic}
}
\maketitle

\abstract{
T-duality and its generalizations are widely recognized either as symmetries or solution-generating techniques in string theory. Recently introduced Jacobi--Lie T-plurality is based on Leibniz algebras whose structure constants ${f_{ab}}^c, {f_c}^{ab}, Z_a, Z^a$ satisfy further conditions. Low dimensional Jacobi--Lie bialgebras were classified a few years ago. We study four- and six-dimensional algebras with structure constants ${f_b}^{ba} = Z^a = 0$ and show that there are several classes consisting of mutually isomorphic algebras.

Using isomorphisms between Jacobi--Lie bialgebras we investigate three- and four-dimensional sigma models related by Jacobi--Lie T-plurality with and without spectators. In the Double Field Theory formulation constant generalized fluxes $F_A$ are used in the literature to transform dilaton field. We extend the procedure to non-constant fluxes and verify that obtained backgrounds and dilatons solve Supergravity Equations. Most of the resulting backgrounds have vanishing curvature scalars and, as can be seen by finding Brinkmann coordinates, represent plane-parallel waves solving Supergravity Equations.
}

\tableofcontents


\section{Introduction}

Gravitational plane-parallel (pp-) waves were repeatedly studied in general relativity and appear frequently in context of string theory as Penrose limits of AdS$\times S$ spacetimes, Dp-brane backgrounds and cosmological models \cite{bfp,prt,blpt}. They also provide exact solutions to string beta function equations \cite{ts}. A few years ago pp-waves were found as duals of flat Minkowski background using \pl\ T-duality with respect to subgroups of Poincare group \cite{HP1,HPP,HPP3}. Recently a new form of duality/plurality, namely \jltpy, has appeared  \cite{rezaseph,melsaka}. In this paper we show that various pp-waves can be constructed by this new form of duality/plurality as well. Moreover, we show that \jltpy\ generates solutions of \sugra s.

Nonlinear sigma models satisfying supplementary conditions are used in string theory to model behavior of strings and $p$-branes propagating in possibly curved backgrounds. They are given by metric $\cG$, Kalb--Ramond field $\cB$ and dilaton $\Phi$. To define a consistent conformally invariant quantum field theory the background bosonic fields in the NS-NS sector have to satisfy one-loop beta function equations \cite{fratsey, calmar} that read
\begin{align}
\label{betaG}
0 &= R_{\mu\nu}-\frac{1}{4}H_{\mu\rho\sigma}H_{\nu}^{\ \rho\sigma}+2\nabla_{\mu}\nabla_\nu\Phi,\\
\label{betaB}
0 &=-\frac{1}{2}\nabla^{\rho}H_{\rho\mu\nu}+\nabla^{\rho}\Phi \, H_{\rho\mu\nu},\\
\label{betaPhi}
0 &= R-\frac{1}{12}H_{\rho\sigma\tau}H^{\rho\sigma\tau}+4\nabla_{\mu}\nabla^\mu\Phi-4\nabla_\mu\Phi \nabla^\mu\Phi .
\end{align}
Here $\nabla$ denotes covariant derivative, $R_{\mu\nu}$ and $R$ are Ricci tensor and scalar curvature of the metric $\cG$ and $H = \mathrm{d} \cB$ is torsion. Equations \eqref{betaG}--\eqref{betaPhi} are also known as Supergravity (SUGRA) Equations.

Crucial insights in string theory were obtained after discovery of dualities. Target space duality (T-duality) relates sigma models in backgrounds with dramatically different curvatures and torsions. Abelian T-duality of bosonic fields \cite{buscher:ssbfe} has been studied thoroughly \cite{recver, aagbl}, extended to fermionic fields \cite{berghull, hassan, bermal} and is understood as an $O(D,D)$ \tfn. To make the $O(D,D)$ group action manifest, authors of Refs. \cite{hullzw, hohuzw} introduced the framework of Double Field Theory (DFT), where transformations of background fields can be found from transformations of the so called generalized metric
\begin{equation}
 \cH_{MN} \equiv \begin{pmatrix} \cG_{mn} -\cB_{mp}\,\cG^{pq}\,\cB_{qn} & \cB_{mp}\,\cG^{pn} \\ -\cG^{mp}\,\cB_{pn} & \cG^{mn}  \end{pmatrix} 
\label{HMN}
\end{equation}
on a manifold of doubled dimension with coordinates $x^M=(x^m, \tilde x_m)$.

Non-Abelian extensions \cite{delaossa:1992vc,aagl} of T-duality proved to be intricate. Introducing \pltd y authors of Refs. \cite{klise,klim:proc} resolved many issues by identifying Lie bialgebra $\cd=(\cg|\tcg)$ and related \dd\ $\D = (\G|\tG)$ as the underlying algebraic structure allowing to construct mutually dual or plural \cite{unge:pltp} sigma models. It was observed \cite{grv, aagl} that dualization with respect to groups $\G$ whose algebra $\cg$ has non-vanishing trace of structure constants ${f_{ab}}^{b} \neq 0$ produces \bkg s that do not satisfy \usugra s. Instead, duality preserves Generalized Supergravity Equations \cite{sasayo,hkc,hlape:bianchisugra} following from $\kappa$-symmetry \cite{sugra2, Wulff:2016tju}. However, in practice solutions to \gsugra s can be sometimes transformed to solutions of equations \eqref{betaG}--\eqref{betaPhi}, see Refs. \cite{wulff, LHsymm}. In recent papers T-duality is often discussed in terms of DFT \cite{dehath, hass}, where the long-standing problem of dependence of dual dilaton on original coordinates can be eliminated at the cost of introduction of Killing vector $\J$ appearing in generalized SUGRA Equations \cite{saka2,HP}.

Recently introduced \JL\ T-duality \cite{rezaseph} generalizes \pltd y by taking \JL\ bialgebras $((\cg,\phi_0),(\tcg,X_0))$ as starting point for construction of related sigma models \cite{rezaseph:H4,ahkr}. \JL\ bialgebras are in one-to-one correspondence to Leibniz algebras \DDp\ considered in Ref. \cite{melsaka}, where \jltpy\ was given in terms of DFT. Authors of Ref. \cite{melsaka} distinguish between three types of Leibniz algebras. Type 3 algebras with $X_0 \neq 0$ produce sigma model backgrounds that depend on dual coordinates and can be hardly considered solutions of (generalized) Supergravity Equations. Isomorphisms of six-dimensional Type 2 algebras, i.e. those with $X_{0}= 2\, Z^a T_a = 0$ and ${f_b}^{ba}\neq 0$, were used in \cite{hlape:JLgsugra} to find \JL\ sigma models satisfying generalized SUGRA Equations. 

In this paper we shall consider Type 1 algebras, i.e. those with $X_{0}=0,\ f_b{}^{ba}= 0$, which are supposed to give sigma models satisfying SUGRA Equations. Given the classification of four- and six-dimensional \JL\ bialgebras found in Ref. \cite{rezaseph:class} we shall investigate isomorphisms of low-dimensional Type 1 algebras and use it to construct mutually plural three- and four-dimensional \JL\ models. Our results confirm that \jltpy\ transforms solution of Eqns. \eqref{betaG}--\eqref{betaPhi} to another solution. In the DFT approach the plural dilaton is found through generalized fluxes that are assumed to be constant in the literature \cite{melsaka,saka2}. We show that the procedure can be extended to non-constant fluxes while still obtaining plural dilaton satisfying Supergravity Equations. 

Many backgrounds obtained this way are interesting from physical point of view, as they are in fact plane-parallel waves. The fact that pp-waves provide exact solutions to string beta function equations \cite{ts} might be interesting in the study of T-duality in presence of higher order $\alpha'$-corrections of the string effective action, see e.g. \cite{marnun,borwulff,hrowuza}.

The paper is organized as follows. In Section \ref{sec:iso} we review properties of \DDp\ algebras and find classes of isomorphic two- and three-dimensional algebras. List of particular isomorphisms can be found in the Appendix. In Section \ref{sec:JLmodels} we summarize the construction of \JL\ T-plural models and plurality transformation. In Section \ref{examples} we present examples of three-dimensional sigma models and atomic \JL\ plurality. Examples of four-dimensional models and plurality with spectators are presented in Section \ref{sec:spec}.

\section{Isomorphisms of low-dimensional algebras}\label{sec:iso}

The $2D$-dimensional DD$^+$ Leibniz algebras were introduced in \cite{melsaka}. Denoting generators as $T_A = (T_a, T^a)$, $a = 1,\ldots,D$; $A = 1,\ldots,2D$, product in  DD$^+$ can be defined using structure coefficients $X_{AB}{}^C$ as
\begin{equation}
T_A \circ T_B = X_{AB}{}^C \, T_C. \label{dd_st_const}
\end{equation}
Besides that one requires that a symmetric bilinear form $\langle , \rangle$ is introduced on DD$^+$, such that there are two subalgebras $\cg$ and $\tcg$ that are maximally isotropic with respect to the form. The generators can be chosen such that
\begin{equation}\label{eta}
\langle T_A, T_B \rangle = \eta_{AB}, 
\qquad 
\eta_{AB} = \begin{pmatrix} 0 & \delta_a^b \\ \delta_a^b & 0 \end{pmatrix}.
\end{equation}
In terms of $T_a$ and  $T^a$ that generate algebras $\cg$ and $\tcg$ the product on DD$^+$ reads
\begin{align}
\begin{split}
 T_{a}\circ T_{b} &= f_{ab}{}^c\,T_{c}\,, \qquad
 T^a\circ T^b = f_c{}^{ab}\,T^c \,,
\\
 T_{a}\circ T^b &= \bigl(f_a{}^{bc} + 2\,\delta_a^b\,Z^c -2\,\delta_a^c\,Z^b\bigr)\,T_c - f_{ac}{}^b\,T^c +2\,Z_a\,T^b\,,
\\
 T^a\circ T_b &= - f_b{}^{ac}\,T_{c} +2\,Z^a\,T_b + \bigl(f_{bc}{}^a +2\,\delta^a_b\,Z_c -2\,\delta^a_c\,Z_b\bigr)\,T^c\,.
\end{split}
\label{eq:DD+components}
\end{align}
Structure constants ${f_{ab}}^c, {f_c}^{ab}, Z_a, Z^a$ satisfy conditions following from Leibniz identities on DD$^+$ and the fact that $\cg$ and $\tcg$ are supposed to be Lie algebras.

There is one-to-one correspondence between Leibniz algebras DD$^+$ and Jacobi-Lie bialgebras $((\cg,\phi_0),(\tcg,X_0))$ studied in \cite{rezaseph,rezaseph:class}, and we can use the classification of four- and six-dimensional Jacobi-Lie bialgebras given in \cite{rezaseph:class}.  As mentioned in the Introduction, non-vanishing $X_{0}= 2\, Z^a T_a$ leads to sigma models with \bkg s depending on dual coordinates \cite{melsaka}. Therefore, for construction of \JL\ models we can use only algebras given here in Tables \ref{Table3} and \ref{tab:algebras} that are duals of Tables 5 and 7 presented in Ref. \cite{rezaseph:class}. The commutation relations and cocycles of \cite{rezaseph,rezaseph:class} are restored by setting $[T_A, T_B]:=\frac{1}{2}(T_A \circ T_B - T_B \circ T_A)$ and $\phi_0 = \beta_a T^a := 2 Z_a T^a$, $X_0 = \alpha^a T_a := 2 Z^a T_a$.

We shall see in the following that several algebras listed in Tables \ref{Table3} and \ref{tab:algebras} are isomorphic in the sense that there are matrices $C$ that transform algebraic relations of one algebra (generated by $T_A$) to those of another algebra (generated by $\hat T_A$) through
\begin{equation}\label{Cplurality}
\hat T_A = C_A{}^B \, T_B , \quad \hat T_A\circ \hat T_B=\hat X_{AB}{}^C \,\hat T_C.
\end{equation}
Since \eqref{eta} has to hold for $\hat T_A$ as well, the conditions on $C$ are
\begin{equation} \label{tfnX}
C_{A}{}^{F}C_{B}{}^{G} X_{FG}{}^H = \hat X_{AB}{}^D\,C_{D}{}^{H}, \qquad C_{A}{}^{F}C_{B}{}^{G} \eta_{FG}=\eta_{AB}.
\end{equation}
We call algebras given by $X_{FG}{}^H$ and $\hat X_{AB}{}^D$ isomorphic or equivalent if there is  a matrix $C$ that solves \eqref{tfnX}. The matrices $C$ are given up to automorphisms of the algebras $\cg$ and $\tcg$. 

\subsection{Type 1 four-dimensional \JL\ algebras}

Four-dimensional \JL\ bialgebras $((\cg,\phi_0),(\tcg,X_0))$ were classified in \cite{rezaseph:class}. Their duals with $X_0 = 0$ needed for construction of \JL\ models are given in the Table \ref{Table3}. To shorten the notation, we denote the bialgebra $((I,T^{1}+T^{2}),(I,0))$ formed by two-dimensional Abelian algebras $\cg$, $\tcg$ as $[1\,|\,1]$. Similarly $[2;\alpha\,|\,1]$ is given by two-dimensional algebra $\cg$ with $T_1 \circ T_2 = T_1$, Abelian $\tcg$ and $\phi_{0} = 2\, Z_a T^a = \alpha\, T^{2}$.

\begin{table}
\begin{center}  
\begin{tabular}{c c l l l lp{0.15mm} }
\multicolumn{5}{l}{}\\
\hline
\hline
{\footnotesize Name of $\cd$} &{\footnotesize ${\tcg}$ }& {\footnotesize ${\cg}$}
&{\footnotesize Product definitions of ${\cg}$}&{\footnotesize $\phi_{0} = 2\, Z_a T^a$}&{\footnotesize Comments} \\
\hline
\vspace{2mm}
{\footnotesize $[1\,|\,1]$}&{\footnotesize $I$}&{\footnotesize $I$}&{\footnotesize $T_i \circ T_j=0$}&{\footnotesize $T^{1}+T^{2}$}&\\
\vspace{2mm}{\footnotesize $[2;\alpha\,|\,1]$}&
{\footnotesize $I$}&{\footnotesize $II$}&{\footnotesize $T_1 \circ T_2 = T_1$}&{\footnotesize $\alpha\, T^{2}$}&{\footnotesize $\alpha\in\real\smallsetminus\{0\}$}\\
\hline
\end{tabular}
\caption{Real four-dimensional Leibniz algebras with $X_0 = 0$ as duals of four-dimensional Jacobi-Lie bialgebras presented in \cite{rezaseph:class}, Table 5. \label{Table3}
}
\end{center} 
\end{table}

There are two isomorphisms among these algebras, namely
\begin{align}
[1|1]\cong& [2;-1|1],\\
[2;\alpha|1]\cong &[2;-\frac{\alpha}{\alpha+1}|1].
\end{align}
Corresponding transformation matrices solving \eqref{tfnX} contain several parameters. We can choose the $C$ matrices e.g. in the form
\begin{equation}
\label{c1121}C= \left(
\begin{array}{cccc}
 0 & 0 & 0 & 1 \\
 -1 & 0 & 0 & 0 \\
 -1 & 1 & 0 & 0 \\
 0 & 0 & -1 & -1 \\
\end{array}
\right)
\end{equation} 
and
\begin{equation}
\label{c2121}C= \left(
\begin{array}{cccc}
 0 & 0 & 1 & 0 \\
 0 & -\frac{1}{\alpha +1} & 0 & 0 \\
 1 & 0 & 0 & 0 \\
 0 & 0 & 0 & -\alpha -1 \\
\end{array}
\right)
\end{equation}
respectively.

\subsection{Type 1 six-dimensional \JL\ algebras}

Six-dimensional \JL\ algebras were also classified in \cite{rezaseph:class}. As mentioned in the Introduction we are interested in models generated by \JL\ algebras with ${f_b}^{ba} = Z^a = 0$. They are displayed in Table \ref{tab:algebras}. The notation follows the same logic as for four-dimensional algebras with numbers in the second and third column refering to Bianchi classification of three-dimensional Lie algebras.

\begin{table}
\begin{center}  
\begin{tabular}{c c l l l lp{0.15mm} }
\multicolumn{5}{l}{}\\
\hline
\hline
{\footnotesize Name of $\cd$} &{\footnotesize ${\tcg}$ }& {\footnotesize ${\cg}$}
&{\footnotesize Product definitions of ${\cg}$}&{\footnotesize $\phi_{0} = 2\, Z_a T^a$}&{\footnotesize Comments} \\
\hline
\vspace{2mm}
{\footnotesize $\{1\,|\,1\}$}&{\footnotesize $I$}&{\footnotesize $I$}&{\footnotesize $T_i \circ T_j =0$}&{\footnotesize $T^{1}$}&\\
\vspace{2mm}{\footnotesize $\{2\,|\,1\}$}&
{\footnotesize $I$}&{\footnotesize $II$}&{\footnotesize $T_2 \circ T_3=T_1$}&{\footnotesize $T^{3}$}&\\
\vspace{2mm}{\footnotesize $\{3\,|\,1\}$}&
{\footnotesize $I$}&{\footnotesize {$III$}}&{\footnotesize {$T_1 \circ T_2=-(T_2+T_3),\ T_1 \circ T_3=-(T_2+T_3)$}}&{\footnotesize $T^{3}-T^{2}$}&\\
\vspace{2mm}{\footnotesize $\{3;b\,|\,1\}$}&
{\footnotesize $I$}&{\footnotesize $III$}&{\footnotesize $T_1 \circ T_2=-(T_2+T_3),\ T_1 \circ T_3=-(T_2+T_3)$}& {\footnotesize $b\,T^{1}$}&{\footnotesize $b\in {\real\smallsetminus\{0\}}$}\\
\vspace{2mm}{\footnotesize $\{4;b\,|\,1\}$}&
{\footnotesize $I$}&{\footnotesize $IV$}&{\footnotesize $T_1 \circ T_2=-(T_2-T_3),\ T_1 \circ T_3=-T_3$}& {\footnotesize $b\,T^{1}$}&{\footnotesize $b\in \real\smallsetminus\{0\}$}\\
\vspace{2mm}{\footnotesize $\{5;b\,|\,1\}$}&
{\footnotesize $I$}&{\footnotesize $V$}&{\footnotesize $T_1 \circ T_2=-T_2,\ T_1 \circ T_3=-T_3$}& {\footnotesize $b\,T^{1}$}&{\footnotesize $b\in \real\smallsetminus\{0\}$}\\
\vspace{2mm}{\footnotesize $\{6_0;b\,|\,1\}$}&
{\footnotesize $I$}&{\footnotesize $VI_{0}$}&{\footnotesize $T_1 \circ T_3=T_2,\ T_2 \circ T_3=T_1$}& {\footnotesize $b\,T^{3}$}&{\footnotesize $b>0$}\\
{\footnotesize $\{6_a;b\,|\,1\}$}&{\footnotesize $I$}&{\footnotesize $VI_{a}$}&{\footnotesize $T_1 \circ T_2=-(aT_2+T_3),\ T_1 \circ T_3=-(T_2+aT_3)$}& {\footnotesize $b\,T^{1}$}&{\footnotesize $a>0,a\neq1$}\\
&&&&&{\footnotesize $b\in \real\smallsetminus\{0\}$}\\
\vspace{2mm}{\footnotesize $\{7_0;b\,|\,1\}$}&
{\footnotesize $I$}&{\footnotesize $VII_{0}$}&{\footnotesize $T_1 \circ T_3=-T_2\, T_2 \circ T_3=T_1$}& {\footnotesize $b\,T^{3}$}&{\footnotesize $b>0$}\\
{\footnotesize $\{7_a;b\,|\,1\}$}&{\footnotesize $I$}&{\footnotesize $VII_{a}$}&{\footnotesize $T_1 \circ T_2=-(aT_2-T_3),\ T_1 \circ T_3=-(T_2+aT_3)$}& {\footnotesize $b\,T^{1}$}&{\footnotesize $a>0$}\\
&&&&&{\footnotesize $b\in \real\smallsetminus\{0\}$}\\
\vspace{2mm}{\footnotesize $\{1\,|\,2\}$}&
{\footnotesize $II$}&{\footnotesize $I$}&{\footnotesize $T_i \circ T_j=0$}&{\footnotesize $T^{1}$}&\\
\vspace{2mm}{\footnotesize $\{2.i\,|\,2\}$}&
{\footnotesize $II$}&{\footnotesize $II.i$}&{\footnotesize $T_1 \circ T_3=T_2$}&{\footnotesize $T^{1}$}&\\
\vspace{2mm}{\footnotesize $\{2.ii\,|\,2\}$}&
{\footnotesize $II$}&{\footnotesize $II.ii$}&{\footnotesize $T_1 \circ T_3=-T_2$}&{\footnotesize $T^{1}$}&\\
\vspace{2mm}{\footnotesize $\{3;b\,|\,2\}$}&
{\footnotesize $II$}&{\footnotesize $III$}&{\footnotesize $T_1 \circ T_2=-(T_2+T_3),\ T_1 \circ T_3=-(T_2+T_3)$}&{\footnotesize $b\,T^{1}$}&{\footnotesize $b\in \real\smallsetminus\{0\}$}\\
\vspace{2mm}{\footnotesize $\{4;b\,|\,2\}$}&
{\footnotesize $II$}&{\footnotesize $IV$}&{\footnotesize $T_1 \circ T_2=-(T_2-T_3),\ T_1 \circ T_3=-T_3$}&{\footnotesize $b\,T^{1}$}&{\footnotesize $b\in \real\smallsetminus\{0\}$}\\
\vspace{2mm}{\footnotesize $\{4.iii;b\,|\,2\}$}&
{\footnotesize $II$}&{\footnotesize $IV.iii$}&{\footnotesize $T_1 \circ T_2=T_2-T_3,\ T_1 \circ T_3=T_3$}&{\footnotesize $b\,T^{1}$}&{\footnotesize $b\in \real\smallsetminus\{0\}$}\\
\vspace{2mm}{\footnotesize $\{5;b\,|\,2\}$}&
{\footnotesize $II$}&{\footnotesize $V$}&{\footnotesize $T_1 \circ T_2=-T_2,\ T_1 \circ T_3=-T_3$}&{\footnotesize $b\,T^{1}$}&{\footnotesize $b\in \real\smallsetminus\{0\}$}\\
\vspace{2mm}{\footnotesize $\{6_0.iii;b\,|\,2\}$}&
{\footnotesize $II$}&{\footnotesize $VI_{0}.iii$}&{\footnotesize $T_1 \circ T_2=T_3,\ T_1 \circ T_3=T_2$}&{\footnotesize $b\,T^{1}$}&{\footnotesize $b >0$}\\
{\footnotesize $\{6_a;b\,|\,2\}$}&{\footnotesize $II$}&{\footnotesize $VI_{a}$}&{\footnotesize $T_1 \circ T_2=-(aT_2+T_3),\ T_1 \circ T_3=-(T_2+aT_3)$}&{\footnotesize $b\, T^1$}& {\footnotesize $a>0,a\neq1$}\\
&&&&&{\footnotesize $b \in \real\smallsetminus\{0\}$}\\
\vspace{2mm}{\footnotesize $\{7_0.i;b\,|\,2\}$}&
{\footnotesize $II$}&{\footnotesize $VII_{0}.i$}&{\footnotesize $T_1 \circ T_2=-T_3,\ T_1 \circ T_3=T_2$}& {\footnotesize $b\,T^{1}$}&{\footnotesize $b>0$}\\
\vspace{2mm}{\footnotesize $\{7_0.ii;b\,|\,2\}$}&
{\footnotesize $II$}&{\footnotesize $VII_{0}.ii$}&{\footnotesize $T_1 \circ T_2=T_3,\ T_1 \circ T_3=-T_2$}& {\footnotesize $b\,T^{1}$}&{\footnotesize $b>0$}\\
{\footnotesize $\{7_a;b\,|\,2\}$}&{\footnotesize $II$}&{\footnotesize $VII_{a}$}&{\footnotesize $T_1 \circ T_2=-(aT_2-T_3),\ T_1 \circ T_3=-(T_2+aT_3)$}&{\footnotesize $b\,T^1$}& {\footnotesize$a > 0$}\\
&&&&&{\footnotesize $b \in \real\smallsetminus\{0\}$}\\
\hline
\end{tabular}
\caption{Real six-dimensional Leibniz algebras with $f_b{}^{ba}=0$, $X_{0}=0$. \label{tab:algebras} These algebras are duals of Jacobi--Lie bialgebras given in \cite{rezaseph:class}, Table 7. Type 2 algebras studied in \cite{hlape:JLgsugra} are not displayed. Second and third column refer to Bianchi classification of three-dimensional algebras. Non-Abelian algebras $\tcg$ are defined by $T^2 \circ T^3 = T^1$.}
\end{center}
\end{table}


Before looking for isomorphisms it is suitable to classify invariants of  DD$^+$ 
like e.g. dimensions of derived algebras given in the Table \ref{tab1}.   

\newpage
\begin{table}[t]
\renewcommand{\arraystretch}{1.3}
\begin{tabular}{|c|c|c|c|l|}
\hline
 Dim. of $\cd^1=$ & Dim. of   & Dim. of          & Type 1 algebras \\
$=\cd_1=\cd\circ\cd $ & $\cd\circ{\cd}^1,{\cd}^1 \circ\cd $ & $\cd_1\circ\cd_1$ & $\cd$   \\ \hline \hline
 5           & 5,5               & 1              &  $ \{4;b|1\},\{4;b|2\},\ b\neq -1 $,\\ 
          &  &  &  $  \{5;b|1\},\{5;b|2\},\ b\neq -1 $,   \\ 
          &  &  & $\{6_0;b|1\}$, $\{6_0.iii;b|2\},\ b\neq 1 $,   \\ 
          &  &  &$\{6_a;b|1\}$, $ \{6_a;b\,|\,2\},\ b\neq 1-a,  $  \\  
          &  &  & $\{7_0;b\,|\,1\} , \{7_0.i;b\,|\,2\}, \{7_0.ii;b\,|\,2\},  $  \\  
          &  &  & $\{7_a;b\,|\,1\} , \{7_a;b\,|\,2\} $  \\ 
          \hline
 4           & 4,4               & 1               & $\{3;b|1\},\{3;b|2\},\ b\neq -2$, \\          &  &  & $\{6_0;1|1\},\{6_0.iii;1|2\},$  \\ 
          &  &  &  $\{6_a;1-a|1\},\{6_a;1-a|2\}$   \\ 
 \cline{3-4} &  & 2  & $\{3|1\}$\\ 
 \cline{2-4} & 3,3 & 1 & $\{2|1\},\{2.i\,|\,2\},\{2.ii\,|\,2\},\{3;-2\,|\,2\} $,  \\
          &             &                   & $\{4;-1\,|\,1\},\{4;-1\,|\,2\},\{4iii;1\,|\,2\}$  \\ \hline
 3           & 3,3              & 0 & $ \{1\,|\,1\}, \{1\,|\,2\}, \{3;-2\,|\,1\}$,  \\ 
           &             &                   & $\{5;-1\,|\,1\},\{5;-1\,|\,2\}$  \\ 
\hline
\end{tabular}
\caption{Invariants of Type 1 six-dimensional algebras.  \label{tab1}}
\end{table}

In conformity with the table of invariants we were able to identify following classes of  isomorphic Type 1 algebras 
\begin{equation}\label{1|1}
\{1\,|\,1\}
\cong\{3;-2\,|\,1\} \cong\{5;-1\,|\,1\}
 \cong\{1\,|\,2\} \cong\{5;-1\,|\,2\}.\end{equation}
\begin{align}\label{2|1}
\{2\,|\,1\}
&\cong\{4;-1\,|\,1\}\cong\{2.i\,|\,2\}\cong\{2.ii\,|\,2\} \cong\{3;-2\,|\,2\}\cong\\ &\cong\{4;-1\,|\,2\}\cong\{4iii;1\,|\,2\}.\nonumber
\end{align}
\begin{align}\label{31|1}
\{3;-1|1\}\cong\{3;2|1\}\cong\{6_0;1|1\}\cong\{6_0.iii;1|2\} ,
\end{align}
\begin{align}\label{3b|1}
 b\neq -2,-1,2 : \  \{3;b|1\}&\cong\{3;\frac{-2\,b}{2+b}|1\}\cong\{3;b|2\}\cong\{3;\frac{-2\,b}{2+b}|2\}\cong
\\
\cong &\{6_{b+1};-b|1\}\cong\{6_{\frac{2-b}{2+b}};\frac{2b}{2+b}|1\} \nonumber
\end{align}
\begin{align}\label{4b|1} b\neq -2,-1 :  \{4;b|1\}\cong\{4;\frac{-b}{1+b}|1\}\cong\{4;b|2\}\cong\{4;\frac{-b}{1+b}|2\} 
\end{align}
\begin{align}\label{5b|1} b\neq -2,-1 :  \{5;b|1\}\cong\{5;\frac{-b}{1+b}|1\}\cong\{5;b|2\}\cong\{5;\frac{-b}{1+b}|2\} 
\end{align}
\begin{align}\label{60b1} b\neq 1 :\{6_0;b|1\}\cong \{6_0.iii2;b|2\}, \\ \nonumber
\{7_0;b|1\}\cong \{7_0.i;b|2\},\cong \{7_0.ii;b|2\}
\end{align}
The matrices $C$ depend again on free parameters. Particular choices are given in Appendix.

Algebras $\{4;-2|1\},\ \{5;-2|1\},\ \{4;-2|2\},\ \{5;-2|2\}$ are not equivalent to any other algebras in the list.

Unfortunately we were not able to find equivalences among algebras $\{6_a;b|1\}$, $ \{6_a;b\,|\,2\}, $ $\{7_a;b\,|\,1\} , \{7_a;b\,|\,2\} $ for general $a \in \mathbb{R}$.

\section{\JL\ models and \jltpy }\label{sec:JLmodels}

The construction of \JL\ symmetric models was described in \cite{melsaka} so here we just summarize necessary formulas. 

In the absence of spectator fields the \bkg\ fields are given by constant matrix $E_0$ and structure of Leibniz algebra DD$^+$. We parametrize elements $g$ of the Lie group $\G$  corresponding to Lie algebra $\cg$ as 
$$g = e^{x^1 T_1} e^{x^2 T_2} \ldots e^{x^D T_D} \in \G.
$$
Matrix $M_A{}^B$ defined via group action
\begin{equation*}
g\-1 \triangleright T_A = M_A{}^B T_B
\end{equation*}
then has the form
\begin{equation}\label{matMAB}
M_A{}^B = 
\begin{pmatrix} 
a_a^b & 0 \\
-\pi^{ac}\,a_c^b & \Exp{-2\Delta} (a\-1)^b_a
\end{pmatrix}.
\end{equation}
The metric $\cG$ and the $\cB$-field of \JL\ model on Lie group $\G$ can be expressed as 
\begin{align}\label{JLmtz}
 \cF_{mn} \equiv \cG_{mn}+\cB_{mn} = \Exp{-2\omega} R_{ab}\,r_m^a\,r_n^b ,\quad
 (R_{ab}) \equiv ((E_0{}\-1){}^{ab}+\pi^{ab})^{-1}\,
\end{align}
where $r_m^a$ are components of right-invariant one-form $\mathrm{d}g g\-1,\ g \in \G$. 
The factor
$$
\Exp{-2\omega} := \Exp{-2\Delta}\tilde{\sigma}
$$
may in general depend on dual coordinates $\tilde x_m$, see Ref. \cite{melsaka}, but for algebras of Type 1 (and Type 2 as well) where $Z^a=0$ we can choose $\tilde{\sigma}=1$.

Denoting components of left-invariant one-form $g\-1\mathrm{d}g$ as $l_m^a$, the standard dilaton $\Phi$ can be found as
\begin{equation}\label{eq:dilaton-JL0}
\Exp{-2\Phi} =\frac{1}{\sqrt{\abs{\det\cG_{mn}}}}\Exp{-2\varphi} \Exp{-\Delta} \,\abs{\det(l_m^a)}
\end{equation}
Knowledge of \fn\ $\varphi$ is complementary to knowledge of dilaton. $\cG$, $\cB$ and $\Phi$ constructed above define \JL\ model on the group $\G$. We will require them to satisfy equations \eqref{betaG}-\eqref{betaPhi}.

To find background fields of the plural model we can use transformation properties of the generalized metric \eqref{HMN} formed from $\cG$ and $\cB$. Denoting $(2D\times 2D)$-matrix
\begin{align}
 \cE_{A}{}^{M}(x)=  \begin{pmatrix} \Exp{\omega} e_a{}^m & 0 \\ -\pi^{ac}\, \Exp{\omega} e_c{}^m & \Exp{-\omega} r^a{}_m \end{pmatrix} , \qquad e_a{}^m r_m{}^b=\delta_a{}^b,
\end{align}
and it's inverse $\cE_{M}{}^{A}$, 
we decompose the generalized metric of \JL\ model as
\begin{equation}
 \cH_{MN} = \cE_M{}^A(x)\, {\cH}_{AB} \,\cE_N{}^B(x)
\label{HE}
\end{equation}
where $\cH_{AB}$ is a constant $(2D\times 2D)$-matrix. Under plurality \eqref{Cplurality} matrix $\cH_{AB}$ transforms as
\begin{equation}\label{HABhat}
\hat \cH_{AB} = C_A{}^C\, {\cH}_{CD} \, C_B{}^D.
\end{equation}
Background fields $\hat\cG$, $\hat\cB$ of the plural model are then obtained from generalized metric $\hat \cH_{MN}$ calculated from $\hat \cH_{AB}$ and $\hat \cE_{M}{}^{A}$ as in \eqref{HE}.

It is useful to note that in terms of the constant matrix $E_0$ formula \eqref{HABhat} is equivalent to
\begin{equation}\label{E0hat}
\hat E_0=\left(\left(P+ E_0 \cdot R \right)^{-1} \cdot \left(Q+E_0 \cdot S \right)\right)^T,
\end{equation}
where the matrices $P,Q,R,S$ are $(D\times D)$-blocks of the \tfn\ \mtr x $C^{-1}$
$$C^{-1}=\begin{pmatrix}
 P & Q \\
 R & S
\end{pmatrix}, \qquad T_A = (C^{-1})_A{}^B \, \hat T_B. $$
This is similar to \pltp y.

Plural dilaton can be found using generalized fluxes $\cF_A$ associated with generalized vielbein $\cE_{A}{}^{M}$ and DFT dilaton $d$ through 
\begin{align*}
\cF_A \equiv \cW^B{}_{AB} + 2\, \cD_A d  = \cE_A{}^M\, F_M=\Exp{\omega}F_A,
\end{align*}
where
\begin{equation*}
\cW_{ABC} \equiv - \cD_A \cE_B{}^M\, \cE_{MC},\quad \cD_A \equiv \cE_A{}^M\,\partial_M,\quad
\Exp{-2d}=\Exp{-2\Phi}\sqrt{|\det \cG|} \,.
\end{equation*}

For Type 1 algebras with ${f_b}^{ba} = Z^a = 0$ considered in this paper\footnote{See \cite{melsaka} for discussion how $F_M$ needs to be modified for ${f_b}^{ba}\neq 0$ and  \cite{hlape:JLgsugra} for examples of resulting \bkg s satisfying generalized \sugra s.} we have
\begin{align}\label{gradfi}
 \partial_M \varphi= \half F_M .
\end{align}
If the flux $F_A$ is constant, which is the case usually assumed in the literature, it transforms under \JL\ plurality \eqref{Cplurality} as  
\begin{equation}\label{hatFA}
\hat F_A=C_A{}^B F_B.
\end{equation}
The plural dilaton $\hat\Phi$ is then found from $\hat\varphi$ given by
\begin{align*}
 \partial_M \hat\varphi= \half \hat F_M = \half \hat \cE_M{}^A \Exp{\hat\omega} \hat F_A.
\end{align*}

These formulas work only if $F_A$ is constant. In cases when the initial \JL\ model has flux $F_A$ dependent on the group \coor s  $x$, like e.g. in the examples in Sections \ref{sec1|1} or \ref{11specIVO},  the formula \eqref{hatFA} must be modified to 
\begin{equation}\label{hatFAxdep}
\hat F_A(X)=C_A{}^B F_B(C_\cdot X)
\end{equation}
where $X=(x_1,\ldots,x_D)$.
In other words by \jltpy\ flux $F_A$ transforms as covariant vector field.

Extension \JLtp y to cases with $n$ spectators $y^m$ can be done as follows. We introduce coordinates $x^M=(y^m, x^m, \tilde y_m, \tilde x_m)$. The generalized metric \eqref{HMN} can be decomposed into product
\begin{align*}
 \cH_{MN} = \cE_M{}^A(x)\, {\cH}_{AB}(y) \,\cE_N{}^B(x)
\end{align*}
of spectator-dependent ${\cH}_{AB}(y)$ and extended vielbein
\begin{align*}
 \cE_{A}{}^{M}(x)=  \begin{pmatrix} \varepsilon_a{}^m & 0 \\
 -\Pi^{ac}\, \varepsilon_c{}^m & \rho^a{}_m \end{pmatrix} 
\end{align*}
with $(n+D)\times(n+D)$ blocks
$$
\varepsilon_a{}^m=\begin{pmatrix}\unit_n & 0 \\ 0 & \Exp{\omega} e_a{}^m  \end{pmatrix}, \quad \Pi^{ac} =\begin{pmatrix}\nul_n&0 \\ 0&\pi^{ac} \end{pmatrix}, \quad \varepsilon_a{}^m  \rho_m{}^b=\delta_a{}^b. 
$$
The \bkg\ fields $\cG$, $\cB$ and fluxes $\cF_A$ are calculated as earlier.

For spectator dependent matrix $E_0(y)$ the formula for \JL\ plurality reads
\begin{equation}\label{E0spec}
\wh E_0(y)=\left(\left(\mathcal{P}+ E_0(y) \cdot \mathcal{R}\right)^{-1} \cdot \left(\mathcal{Q}+E_0(y) \cdot \mathcal{S}\right)\right)^T,
\end{equation}
where the matrices $\mathcal P,\mathcal Q,\mathcal R,\mathcal S $ are obtained by extension  of the $D\times D$ blocks of the \tfn\ \mtr x $C^{-1}$
$$C^{-1}=\begin{pmatrix}
\mathcal{P}&\mathcal{Q} \\
\mathcal{R} & \mathcal{S}
\end{pmatrix}$$
to $(n+D)\times (n+D)$ matrices
\begin{equation*}
\mathcal{P} =\begin{pmatrix}\unit_n &0 \\ 0&P \end{pmatrix}, \quad \mathcal{Q} =\begin{pmatrix}\nul_n&0 \\ 0&Q \end{pmatrix}, \quad \mathcal{R} =\begin{pmatrix}\nul_n&0 \\ 0&R \end{pmatrix}, \quad \mathcal{S} =\begin{pmatrix}\unit_n &0 \\ 0& S \end{pmatrix}
\end{equation*}
to accommodate the spectator fields.

Beside that we need formula for spectator-dependent dilaton. In \cite{melsaka,saka2} it was suggested in the form
\begin{equation}\label{specdil393}
 \Exp{-2\Phi(x,y)} = \frac{1}{\sqrt{\abs{\det \cG(x,y)}}}\Exp{-2\hat d(y)}\Exp{-2\varphi(x)} \Exp{-\Delta(x)}  \abs{\det \left( l_m{}^a(x)\right)}.
\end{equation}
To get solutions of \sugra s we  choose 
\begin{equation}\label{dilmelsaka}
\hat d(y)=-\frac{1}{4}\ln\frac{(\det E_0(y))^2}{\det E_{0S}(y)}, \quad E_{0S}(y)=\half(E_0(y)+E_0^T(y)).
\end{equation}

\section{Examples of atomic \JLtp y}\label{examples}

To be able to apply \jltpy\ when looking for \JL\ models satisfying \sugra s we must first find a model satisfying these \eqn s  for at least one of the algebras in the equivalence class. It means finding matrix $E_0$ and dilaton for given algebra such that the \JL\ model constructed by procedure described in the preceding section satisfies \sugra s. This may be a difficult task, but in many cases it is tractable.

When analyzing \bkg s obtained in the following sections we often encounter pp-waves. Their metric in the Brinkmann coordinates $(u, v, z_3, z_4, \ldots , z_{m-2})$ is usually written as
\begin{equation}\label{pp-wave}
ds^2 = - K(u,\vec z) du^2 + 2dudv + d{\vec z}^2,
\end{equation}
where $d{\vec z}^2$ is the Euclidean metric in the transversal space with coordinates $(z_3, z_4, \ldots , z_{m-2})$. 
Metric \eqref{pp-wave} also has particularly simple Ricci tensor with only one non-vanishing component
\begin{equation}\label{Ruu}
R_{uu}=\frac{1}{2}\left(\partial_3^2 K + \partial_4^2 K + \ldots + \partial_{m-2}^2 K\right).
\end{equation}
Moreover, any curvature invariants calculated for a pp-wave \bkg\ vanish \cite{prt}. These properties allow us to identify pp-waves in the examples below. More specifically, we will be able to bring many backgrounds to the form \eqref{pp-wave} where
\begin{equation}\label{ppw-K}
K(u,\vec z) = K_i(u) z_i^2.
\end{equation}
For further details on pp-waves and their appearance in string theory see e.g. \cite{bfp,prt,blpt}.

In this Section we present examples of atomic plurality of three-dimensional \JL\ models with initial models satisfying Supergravity Equations.

\subsection{\JL\ models plural to $\{1|1\}$}\label{sec1|1}

Let us consider six-dimensional \JL\ bialgebra $\{1|1\}$ composed of Abelian subalgebras, i.e. ${f_{ab}}^c = {f_c}^{ab} = 0$, and $\phi_0 = 2\, Z_a T^a = T^1$. Then $r^a_m$ is identity matrix, $\pi^{ab}$ vanishes, $\tilde \sigma = 1$ and $\Delta = \omega = \frac{x_1}{2}$. All three-dimensional \JL\ backgrounds \eqref{JLmtz} given by $\{1|1\}$ are conformally flat because $\cf=e^{-x_1}E_0 $ and $E_0$ is a constant matrix. 

General solution of \sugra s is obtained for
\begin{equation}\label{E_0_1|1}
E_0 = \left(
\begin{array}{ccc}
 \lambda_1 & \lambda_2 & \lambda_3 \\
 \lambda_4 & \lambda_5 & \lambda_6 \\
 \lambda_7 & \lambda_6 & \frac{\lambda_6^2}{\lambda_5} \\
\end{array}
\right), \qquad \Phi =-\frac{1}{8}x_1 + c_1 e^{-x_1}+c_2
\end{equation}
where $\lambda_i, c_i \in \mathbb{R}, \lambda_5 \neq 0$, are constants. For any $\lambda_i$ the resulting background has vanishing torsion and curvature scalar. The only non-vanishing component of Ricci tensor is $R_{11}=-\frac{1}{4}$. Since the properties of this model and all plural ones do not depend on constants $\lambda_i$ that can be eliminated by coordinate transformations, we may choose $E_0$ e.g. as
\begin{equation*}
E_0=\left(
\begin{array}{ccc}
 1 & 1 & 0 \\
 1 & 1 & 1 \\
 0 & 1 & 1 \\
\end{array}
\right).
\end{equation*}

The constant shift of dilaton given by $c_2$ in $\Phi$ is irrelevant. However, for nonzero $c_1$ we get non-constant flux 
$$F_A=\left(\frac{3}{4}-2\,c_1 e^{-x_1},0,0,0,0,0\right).$$
Background $\cf=e^{-x_1}E_0$ can be brought by \tfn\ 
$$ x_1=-\ln (u),\ x_2=-v+\half  \ln
   (u)-\frac{z_3^2}{4 u},\ x_3=\frac{z_3 \left(4
   \sqrt{u}+z_3\right)}{4 u}-\frac{\ln
   (u)}{2}+v
$$to the Brinkmann form of plane-parallel wave 
\begin{equation}
ds^2 = -\frac{z_3^2}{4 u^2} du^2 + 2dudv + dz_3^2. \label{ppwave4}
\end{equation} 

Now when we have a solution of \sugra s we can look for plural models. Nevertheless, not all models on isomorphic algebras \eqref{1|1} can be constructed this way. 
Plural model \crspto plurality $\{1\,|\,1\}\cong\{5;-1\,|\,1\}$ given e.g. by matrix
\begin{equation}\label{11to511}
 C= 
\left(
\begin{array}{cccccc}
 -1 & 0 & 0 & 0 & 0 & 0 \\
 0 & 0 & 0 & 0 & 1 & 0 \\
 0 & 0 & 0 & 0 & 0 & 1 \\
 0 & 0 & 0 & -1 & 0 & 0 \\
 0 & 1 & 0 & 0 & 0 & 0 \\
 0 & 0 & 1 & 0 & 0 & 0 \\
\end{array}
\right)
\end{equation}
does not exist because $\left(P+ E_0 \cdot R \right)$ in \eqref{E0hat} is not invertible and matrix $\hat E_0$ does not exist. This holds for general matrix $E_0$ in \eqref{E_0_1|1}.

\subsubsection{Models \crspto plurality $\{1\,|\,1\}\cong\{3;-2\,|\,1\}$}

As we can see in Table \ref{tab:algebras}, \JL\ bialgebra $\{3;-2\,|\,1\}$ is given by relations
$$T_1 \circ T_2 = -(T_2+T_3),\quad T_1 \circ T_3 = -(T_2+T_3), \quad \phi_0 = 2\, Z_a T^a = -2\, T^1.$$
The $C$-matrix of isomorphism $\{1\,|\,1\}\cong\{3;-2\,|\,1\}$ reads e.g.
\begin{equation}\label{11to321}
C= \left(
\begin{array}{cccccc}
 -2 & 0 & 0 & 0 & 0 & 0 \\
 0 & 0 & \frac{1}{2} & 0 & 1 & 0 \\
 0 & 0 & -\frac{1}{2} & 0 & 1 & 0 \\
 0 & 0 & 0 & -\frac{1}{2} & 0 & 0 \\
 0 & \frac{1}{2} & 0 & 0 & 0 & 1 \\
 0 & \frac{1}{2} & 0 & 0 & 0 & -1 \\
\end{array}
\right).
\end{equation}
Matrix $\hat E_0$ calculated via \eqref{E0hat} is
\begin{equation*}
\hat E_0 = \left(
\begin{array}{ccc}
 0 & -1 & -3 \\
 3 & 1 & 2 \\
 1 & 0 & 1 \\
\end{array}
\right).
\end{equation*} 
Since $\pi^{ab}$ vanishes and $\omega = -x_1$, we get background \eqref{JLmtz} of plural \JL\ model in the form
\begin{equation}
\label{321}
\hat\cf= \left(
\begin{array}{ccc}
 0 & e^{2 x_1}-2 & -e^{2 x_1}-2 \\
 e^{2 x_1}+2 & e^{-2 x_1} & e^{-2 x_1}+1
   \\
 2-e^{2 x_1} & e^{-2 x_1}-1 & e^{-2 x_1}
   \\
\end{array}
\right).
\end{equation}
From \eqref{gradfi} and \eqref{hatFAxdep} we find $\hat \varphi = -\frac{3}{4}\,x_1+c_1 e^{2x_1}$ and up to a constant shift the plural dilaton obtained from \eqref{eq:dilaton-JL0} is
$$
\hat\Phi=-\frac{3}{4} \,x_1+c_1 e^{2x_1}.
$$
Dilaton $\hat\Phi$ together with metric and $\cB$-field obtained from \eqref {321} satisfy \sugra s. It is important that $C$-matrix \eqref{11to321} transformed the coordinate $x_1$ appearing in the non-constant flux $F_A$ as $x_1 \mapsto -2 x_1$. In general it is possible that dual coordinates enter $\partial_M \hat\varphi$.

Background \eqref{321} is torsionless with vanishing scalar curvature, and the only non-vanishing component of Ricci tensor is $R_{11}=3$. Using the \tfn\ 

we can find the Brinkmann form of a pp-wave metric
\begin{equation}\label{ppwave34}
ds^2 = \frac{3 z_3^2}{4 u^2} du^2 + 2dudv + dz_3^2. \qquad  
\end{equation}

\subsubsection{Models \crspto plurality $\{1\,|\,1\}\cong\{1\,|\,2\}$}\label{412}

\JL\ T-plurality given e.g. by matrix 
\begin{equation}\label{11to12}
C=\left(
\begin{array}{cccccc}
 1 & 0 & 0 & 0 & 0 & 0 \\
 0 & 1 & 0 & 0 & 0 & 0 \\
 0 & 0 & 1 & 0 & 0 & 0 \\
 0 & 0 & 0 & 1 & 0 & 0 \\
 0 & 0 & -1 & 0 & 1 & 0 \\
 0 & 1 & 0 & 0 & 0 & 1 \\
\end{array}
\right)
\end{equation}
allows to construct sigma model on algebra $\{1\,|\,2\}$, where $T_a \circ T_b = 0$ and the only non-trivial structure constants are found from
$$
T^2 \circ T^3 = T^1, \quad \phi_0 = 2\, Z_a T^a = T^1.
$$
This time we have
\begin{equation*}
\pi^{ab} = \left(
\begin{array}{ccc}
 0 & 0 & 0 \\
 0 & 0 & 1-e^{-x_1} \\
 0 & e^{-x_1}-1 & 0 \\
\end{array}
\right), \quad
\hat E_0 = \left(
\begin{array}{ccc}
 0 & 2 & 1 \\
 0 & 1 & 1 \\
 -1 & 1 & 1 \\
\end{array}
\right), \quad \omega = \frac{x_1}{2} 
\end{equation*}
and we obtain \bkg
$$\hat\cf= \left(
\begin{array}{ccc}
 e^{-3 x_1} \left(e^{2 x_1}-1\right) & e^{-2 x_1} \left(e^{x_1}+1\right)
   & e^{-2 x_1} \\
 e^{-2 x_1} \left(e^{x_1}-1\right) & e^{-x_1} & e^{-x_1} \\
 -e^{-2 x_1} & e^{-x_1} & e^{-x_1} \\
\end{array}
\right).$$
The matrix \eqref{11to321} transforms $x_1 \mapsto x_1$. From \eqref{eq:dilaton-JL0}, \eqref{hatFA}, and \eqref{gradfi} we find $\hat \varphi = -\frac{3}{8}\,x_1+c_1 e^{-x_1}$ and dilaton
$$\hat \Phi=-\frac{1}{8}\, x_1+c_1 e^{-x_1}$$
that together with $\hat\cf$ satisfy \sugra s.

The background has nontrivial $\cB$-field but vanishing torsion, curvature scalar $R=0$, and the only non-vanishing component of Ricci tensor is $R_{11}=-\frac{1}{4}$. By the \tfn\ 
$$
x_1=-\ln \left(u\right),\qquad x_2=-\frac{u^3+4 v u-4 z_3 \sqrt{u}+z_3^2-4 u \ln \left(u\right)}{4 u},$$
$$x_3=\frac{u^3+4 v u+z_3^2}{4 u}$$
we can bring the metric to the pp-wave form \eqref{ppwave4}. Therefore, in this case \jltpy\ acts as a coordinate transformation and gauge transformation changing $\cB$-field.

\subsubsection{Models \crspto plurality $\{1\,|\,1\}\cong\{5;-1\,|\,2\}$}\label{512}

The remaining \JL\ T-plurality transformation in the class \eqref{1|1} is given e.g. by matrix 
\begin{equation}\label{11to512}
C=\left(
\begin{array}{cccccc}
 -1 & 0 & 0 & 0 & 0 & 0 \\
 0 & 0 & 0 & 0 & 1 & 0 \\
 0 & 0 & 0 & 0 & 0 & -1 \\
 0 & 0 & 0 & -1 & 0 & 0 \\
 0 & 1 & 0 & 0 & 0 & 1 \\
 0 & 0 & -1 & 0 & 1 & 0 \\
\end{array}
\right)
\end{equation}
transforming algebra $\{1\,|\,1\}$ to $\{5;-1\,|\,2\}$. Background field
$$\hat\cf=\left(
\begin{array}{ccc}
 e^{x_1}-e^{3 x_1} & -e^{2 x_1} &
   -e^{x_1} \left(e^{x_1}+1\right) \\
 e^{2 x_1} & e^{x_1} & e^{x_1}+1 \\
 e^{x_1} \left(e^{x_1}-1\right) &
   e^{x_1}-1 & e^{x_1} \\
\end{array}
\right)$$
and dilaton $\hat \Phi=\frac{1}{8}\,x_1+c_1 e^{x_1} $ obtained by the plurality \tfn s satisfy \sugra s. The \bkg\ is again torsionless and by \tfn\ 
$$
x_1=\ln \left(u\right),\quad x_2=\frac{u^3+4 v u+4 z_3 \sqrt{u}+z_3^2-2 u \ln \left(u\right)}{4 u},$$
$$x_3=-\frac{u^3+4 v u+z_3^2-2 u \ln \left(u\right)}{4 u}
$$
we can bring the metric to the form of pp-wave \eqref{ppwave4}.

\subsection{\JL\ models plural to $\{3;b|1\},\, b\neq 0, -2, -1, 2$}
 
As another example let us consider \JL\ model on algebra $\{3;b|1\}$ in the class \eqref{3b|1}, where $b \neq 0, -2, -1, 2$. 
Background 
$$
\cf=e^{-(b+2) x_1} \left(
\begin{array}{ccc}
 0 & \frac{1}{2} \left(e^{2 x_1}+1\right) & -\frac{1}{2} \left(e^{2 x_1}-1\right) \\
\frac{1}{2} \left(e^{2 x_1}+1\right) & e^{-2 x_1} & e^{- 2 x_1} \\
 -\frac{1}{2} \left(e^{2 x_1}-1\right) & e^{- 2 x_1} & e^{- 2 x_1} \\
\end{array}
\right)
$$
and dilaton 
$$
\Phi=\left(\frac{2}{b}-\frac{b}{8}\right)x_1
$$
satisfying \sugra s are obtained from
$$
E_0=\left(
\begin{array}{ccc}
 0 & 1 & 0 \\
 1 & 1 & 1 \\
 0 & 1 & 1 \\
\end{array}\right), \quad \varphi = \frac{3 b ^2+8 b +16}{8 b} x_1.
$$

Investigating curvature properties of $\cG = \cF$ we find that scalar curvature vanishes and the only non-vanishing component of Ricci tensor is $R_{11}=4-\frac{b^2}{4}$. Using rather complicated transformation of coordinates 
\begin{align*}
x_1&=-\frac{\ln (u)}{b},\\ x_2&=\frac{1}{8} \left(-\frac{-4 z_3 u^{\frac{1}{2}-\frac{2}{b}}+u^{\frac{b-2}{b}}+8 b u v+2 (b+4)
   z_3^2}{u}-\frac{\ln (u)}{b}\right),\\ x_3&=\frac{1}{8} \left(\frac{4 z_3 u^{\frac{1}{2}-\frac{2}{b}}-u^{\frac{b-2}{b}}+8 b u v+2 (b+4)
   z_3^2}{u}+\frac{\ln (u)}{b}\right)
\end{align*}
we can bring the metric to the Brinkmann form of  pp-wave
\begin{equation}\label{pp_b2_16}
ds^2 = -\frac{\left(b^2-16\right) z_3^2}{4 b ^2 u^2} du^2 + 2dudv + dz_3^2.
\end{equation}
As can be seen, solutions of \sugra s for vanishing dilaton exist only for $b=\pm 4$, i.e., flat metric. 

Plural models cannot be obtained by pluralities  $\{3;b|1\}\cong \{6_{b+1};-b|1\}\cong\{6_{\frac{2-b}{2+b}};\frac{2b}{2+b}|1\}$  because for all matrices $C$ that provide these isomorphisms matrices $\hat E_0$ given by \eqref{E0hat} do not exist.

\subsubsection{Models \crspto plurality $\{3;b|1\}\cong \{3;\frac{-2b}{2+b}\,|\,1\},\ b\neq -2$}

\JL\  T-plurality $\{3;b|1\}\cong \{3;\frac{-2b}{2+b}\,|\,1\}$ is given e.g. by matrix
$$ C=\left(
\begin{array}{cccccc}
 -\frac{2}{b+2} & 0 & 0 & 0 & 0 & 0 \\
 0 & \frac{1}{4} & -\frac{1}{4} & 0 & 1 & 1 \\
 0 & -\frac{1}{4} & \frac{1}{4} & 0 & 1 & 1 \\
 0 & 0 & 0 & -\frac{b}{2}-1 & 0 & 0 \\
 0 & \frac{1}{4} & \frac{1}{4} & 0 & 1 & -1 \\
 0 & \frac{1}{4} & \frac{1}{4} & 0 & -1 & 1 \\
\end{array}
\right).
$$
Transformed \JL\ \bkg\ then is 
$$\hat\cf=\hat \cG=\left(
\begin{array}{ccc}
 -\frac{e^{\frac{2 b x_1}{b+2}}}{(b+2)^2} &
   -\frac{\left(e^{2 x_1}+2\right) e^{-\frac{4
   x_1}{b+2}}}{2 (b+2)} & \frac{\left(e^{2 x_1}-2\right)
   e^{-\frac{4 x_1}{b+2}}}{2 (b+2)} \\
 -\frac{\left(e^{2 x_1}-2\right) e^{-\frac{4 x_1}{b+2}}}{2
   (b+2)} & e^{-\frac{2 (b+4) x_1}{b+2}} & e^{-\frac{2
   (b+4) x_1}{b+2}} \\
 \frac{\left(e^{2 x_1}+2\right) e^{-\frac{4 x_1}{b+2}}}{2
   (b+2)} & e^{-\frac{2 (b+4) x_1}{b+2}} & e^{-\frac{2
   (b+4) x_1}{b+2}} \\
\end{array}
\right). $$
Its torsion and scalar curvature vanishes and the only non-vanishing component of Ricci tensor is $R_{11}=\frac{(b+4) (3 b+4)}{(b+2)^2}$.

\sugra s are satisfied by dilaton
\begin{equation}\label{dilb43b4}
\Phi= -\frac{(b+4) (3 b+4) x_1}{4 b (b+2)}
\end{equation}
obtained from \eqref{hatFA}, \eqref{gradfi}.
We can bring the metric to the pp-wave form
\begin{equation}\label{ppb43b4}
ds^2 = \frac{(b+4) (3 b+4) z_3^2}{4 b^2 u^2} du^2 + 2dudv + dz_3^2.
\end{equation} by the \tfn
\begin{align*}
x_1 & =\frac{(b+2) \ln (u)}{2 b},\\
x_2 & =\frac{z_3 u^{\frac{2}{b}+\frac{3}{2}}+4 b u v- b z_3^2-4 z_3^2}{2 u}+\frac{\ln (u)}{4 b},\\
x_3 & =\frac{z_3 \left(u^{\frac{2}{b}+\frac{3}{2}}+4 z_3\right)+b \left(z_3^2-4 u v\right)}{2 u}-\frac{\ln (u)}{4 b}.
\end{align*}

\subsubsection{Models \crspto plurality $\{3;b|1\}\cong \{3;b|2\},\ b\neq -2$}

\JL\  T-plurality is given e.g. by matrix
\begin{equation}C=\left(
\begin{array}{cccccc}
 1 & 0 & 0 & 0 & 0 & 0 \\
 0 & 1 & 0 & 0 & 0 & 0 \\
 0 & 0 & 1 & 0 & 0 & 0 \\
 0 & 0 & 0 & 1 & 0 & 0 \\
 0 & 0 & -\frac{1}{b+2} & 0 & 1 & 0 \\
 0 & \frac{1}{b+2} & 0 & 0 & 0 & 1 \\
\end{array}
\right).\label{c50}
\end{equation}
Plural \JL\ \bkg\ in the group \coor s has rather complicated form. Nevertheles, together with dilaton
$$
\Phi= \left(\frac{2}{b}-\frac{b}{8}\right)x_1$$ they satisfy \sugra s. The torsion and scalar curvature vanish and in the Brinkmann coordinates $(u, v, z_3)$ given by
\begin{align}
x_1 &=-\frac{\ln (u)}{b},\\
x_2 + x_3 &= \frac{1}{4} \left(4 z_3 u^{-\frac{b+4}{2 b}}-u^{-2/b}\right)\\
x_2 - x_3 &= -\frac{1}{4} \left(\frac{2 u^{\frac{4}{b}+2}}{(b+2)^3}+\frac{2 (b+4) z_3^2}{u}+\frac{\ln (u)}{b}+8 b v\right)
\end{align}
the metric has the form \eqref{pp_b2_16} so that \JL\  T-plurality  given by \eqref{c50} is equivalent to a \coor\ \tfn.
 
\subsubsection{Models \crspto plurality $\{3;b|1\}\cong \{3;\frac{-2\,b}{2+b}|2\},\ b\neq -2$}

\JL\  T-plurality is given e.g. by matrix
$$ C=\left(
\begin{array}{cccccc}
 -\frac{2}{b+2} & 0 & 0 & 0 & 0 & 0 \\
 0 & -\frac{1}{2} & \frac{1}{2} & 0 & -\frac{2}{b+2} &
   -\frac{2}{b+2} \\
 0 & \frac{1}{2} & -\frac{1}{2} & 0 & -\frac{2}{b+2} &
   -\frac{2}{b+2} \\
 0 & 0 & 0 & -\frac{b}{2}-1 & 0 & 0 \\
 0 & -\frac{1}{4} (b+2) & 0 & 0 & 0 & 1 \\
 0 & -\frac{1}{4} (b+2) & 0 & 0 & 0 & -1 \\
\end{array}
\right).
$$
Plural \JL\ \bkg\ in the group \coor s has rather complicated form. Nevertheles, together with dilaton \eqref{dilb43b4}
they satisfy \sugra s. Its torsion and scalar curvature vanishes,$R_{11}=\frac{(b+4) (3 b+4)}{(b+2)^2}$, and the metric can be brought to the Brinkmann form \eqref{ppb43b4}.
   
\begin{table}
\begin{center}
{\renewcommand{\arraystretch}{1.6}
\begin{tabular}{|c || c | c | c |}
\hline
Algebra & Metric & Dilaton  \\
\hline \hline  
$ \{1|1\} $ & $-\frac{1}{4 u^2} z_3^2 du^2 + 2dudv + dz_3^2$ &$\frac{1}{8}\ln \left(u\right)$ \\
\hline
$ \{3;-2|1\} $ & $\frac{3}{4u^2} z_3^2 du^2 + 2dudv + dz_3^2$ & $ -\frac{3}{8}\ln \left(u\right)$ \\
\hline
$ \{1|2\} $ & $-\frac{1}{4 u^2} z_3^2 du^2 + 2dudv + dz_3^2$ & $\frac{1}{8}\ln \left(u\right)$ \\
\hline
$ \{5;-1|2\} $ & $-\frac{1}{4 u^2} z_3^2 du^2 + 2dudv + dz_3^2$ & $\frac{1}{8}\ln \left(u\right)$ \\
\hline
$ \{3;b|1\} $ & $ -\frac{b^2-16}{4 b^2 u^2} z_3^2 du^2 + 2dudv + dz_3^2 $ & $\frac{b^2-16}{8 b^2}\ln \left(u\right)$ \\
\hline
$ \{3;\frac{-2\, b}{2+b}|1\} $ & $ \frac{(b+4)(3\,b+4)}{4b^2 u^2} z_3^2 du^2 + 2dudv + dz_3^2 $ & $-\frac{(b+4) (3 b+4)}{8 b^2} \ln (u)$
\\
\hline
$ \{3;b|2\} $ & $ -\frac{b^2-16}{4 b^2 u^2} z_3^2 du^2 + 2dudv + dz_3^2 $ & $\frac{b^2-16}{8 b^2}\ln \left(u\right)$ \\
\hline
$ \{3;\frac{-2\, b}{2+b}|2\} $ & $ \frac{(b+4)(3\,b+4)}{4b^2 u^2} z_3^2 du^2 + 2dudv + dz_3^2 $ & $-\frac{(b+4) (3 b+4)}{8 b^2} \ln
   (u)$\\
\hline
\end{tabular}
}
\normalsize \normalsize \caption {Brinkmann forms of three dimensional metrics and dilatons of models without spectators. \label{table4}}
\end{center}
\end{table}
   
\section{Examples of \JLtp y with spectators}\label{sec:spec}

In this section we shall investigate four-dimensional \JL\ models and \JL\ T-plurality with one or two spectators. In the presence of spectators it is non-trivial to find initial \JL\ model
satisfying \sugra s. Our strategy is to start with Leibniz algebras DD$^+$ having Abelian subalgebras $\cg, \tcg$ and search for $E_0(y)$ and $\varphi$ giving flat metrics and vanishing dilaton. In the examples with one spectator 
we denote the spectator $ y_1=t$.

\subsection{Plurality with one spectator - plurals to flat model on $\{1|1\}$}\label{11spec}

Four-dimensional flat and torsionless \JL\ model given by the algebra $\{1|1\}$ can be obtained e.g. from
\begin{equation}
E_0(t) = \left(
\begin{array}{cccc}
 0 & t & 0 & 0 \\
 t & 0 & 0 & 0 \\
 0 & 0 & 1 & 0 \\
 0 & 0 & 0 & 1 \\
\end{array}
\right).
\end{equation}
Flat \JL\ \bkg\ then is 
\begin{equation}
\cf(t,x)=\left(
\begin{array}{cccc}
 0 & e^{-\frac{x_1}{2}} t & 0 & 0 \\
 e^{-\frac{x_1}{2}} t & 0 & 0 & 0 \\
 0 & 0 & e^{-x_1} & 0 \\
 0 & 0 & 0 & e^{-x_1} \\
\end{array}
\right)
\end{equation} and usual Supergravity Equations are satisfied for vanishing dilaton $\Phi = 0$ \crspto $\varphi=\half x_1, \hat d = -\half \ln t$.

Having initial \JL\ model satisfying \sugra s we can look for plural models and check if they satisfy these \eqn s as well.

\subsubsection{ \JL\ models corresponding to $\{3;-2|1\}$ }

Formula \eqref{E0spec} for the matrix \eqref{11to321} gives
$$
\hat E_0(t)=\left(
\begin{array}{cccc}
 0 & -2 t & 0 & 0 \\
 -2 t & 0 & 0 & 0 \\
 0 & 0 & \frac{5}{4} & \frac{3}{4} \\
 0 & 0 & \frac{3}{4} & \frac{5}{4} \\
\end{array}
\right).
$$ 
For the algebra $\{3;-2|1\}$ we get curved \bkg\ with vanishing torsion
$$ \hat \cf(t, x) =\left(
\begin{array}{cccc}
 0 & -2 e^{x_1} t & 0 & 0 \\
 -2 e^{x_1} t & 0 & 0 & 0 \\
 0 & 0 & e^{-2 x_1}+\frac{e^{2 x_1}}{4} &
   e^{-2 x_1}-\frac{e^{2 x_1}}{4} \\
 0 & 0 & e^{-2 x_1}-\frac{e^{2 x_1}}{4} &
   e^{-2 x_1}+\frac{e^{2 x_1}}{4} \\
\end{array}
\right) $$
that together with dilaton
$$
\hat\Phi=-x_1
$$
obtained from the formula \eqref{specdil393} satisfy Supergravity Equations.

We were not able to find the Brinkmann form of this metric. Nevertheless, we have checked that curvature scalar $R$, Kretschmann scalar $R_{\mu\nu\kappa\lambda}R^{\mu\nu\kappa\lambda}$ and $R_{\mu\nu}R^{\mu\nu}$ vanish. This complies with the fact that for a pp-wave metric all curvature scalars vanish. We have also checked that the first $\alpha'$ correction to the \sugra s \cite{HT,borsatowulf,iranci} vanishes indicating that $\hat \cf(t, x)$ might be an exact pp-wave solution \cite{ts}.

\subsubsection{\JL\ models corresponding to $\{5;-1|1\}$   }

Formula \eqref{E0spec} for the matrix \eqref{11to511}
gives
$$ \hat E_0(t)=\left(
\begin{array}{cccc}
 0 & -t & 0 & 0 \\
 -t & 0 & 0 & 0 \\
 0 & 0 & 1 & 0 \\
 0 & 0 & 0 & 1 \\
\end{array}
\right). $$ 
From the algebra $\{5;-1|1\}$ we get
$$ \hat \cf(t, x) =\left(
\begin{array}{cccc}
 0 & -e^{\frac{x_1}{2}} t & 0 & 0 \\
 -e^{\frac{x_1}{2}} t & 0 & 0 & 0 \\
 0 & 0 & e^{-x_1} & 0 \\
 0 & 0 & 0 & e^{-x_1} \\
\end{array}
\right) $$
that together with dilaton
$$\hat\Phi=-x_1$$
satisfy Supergravity Equations. Scalar curvature of the metric is zero.

For bringing the metric into the form of a homogeneous pp-wave \cite{prt}
\begin{equation}\label{ppwave51}
ds^2 =\frac{2 \left(z_3^2+z_4^2\right)}{u^2} du^2 + 2dudv + dz_3^2 + dz_4^2  
\end{equation}
we need transformation 
\begin{align*}
t &=-{\sqrt{\frac{-2 u v+z_3^2+z_4^2}{2 u}}}, & x_2 &=u z_4,\\
x_1 &=2 \ln (u), & x_3 &=u z_3.
\end{align*}

\subsubsection{ \JL\ models corresponding to $\{1|2\}$ }

Formula \eqref{E0spec} for the matrix \eqref{11to12} gives
$$
\hat E_0(t)=\left(
\begin{array}{cccc}
 0 & t & 0 & 0 \\
 t & 0 & 0 & 0 \\
 0 & 0 & \frac{1}{2} & \frac{1}{2} \\
 0 & 0 & -\frac{1}{2} & \frac{1}{2} \\
\end{array}
\right).
$$ 
From the algebra $\{1|2\}$ we get \bkg
$$
\hat \cf(t, x) =\left(
\begin{array}{cccc}
 0 & e^{-\frac{x_1}{2}} t & 0 & 0 \\
 e^{-\frac{x_1}{2}} t & 0 & 0 & 0 \\
 0 & 0 & \frac{e^{x_1}}{e^{2 x_1}+1} &
   \frac{1}{e^{2 x_1}+1} \\
 0 & 0 & -\frac{1}{e^{2 x_1}+1} &
   \frac{e^{x_1}}{e^{2 x_1}+1} \\
\end{array}
\right)
$$
that together with  dilaton
$$
\hat\Phi=\frac{1}{4} \ln \left(\frac{e^{4 x_1}}{\left(e^{2 x_1}+1\right){}^2}\right)
$$
obtained from the formula \eqref{specdil393} satisfy Supergravity Equations. Scalar curvature vanishes.
For bringing the metric into the Brinkman form
\begin{equation}\label{ppwave12}
ds^2 =\frac{2 u^2 \left(u^4-5\right) \left(z_3^2+z_4^2\right)}{\left(u^4+1\right)^2} du^2 + 2dudv + dz_3^2 + dz_4^2  
\end{equation}
we need transformation
\begin{align*}
t &=\sqrt{\frac{u^4 \left(z_3^2+z_4^2\right)-2 u^5 v-2 u v-z_3^2-z_4^2}{2(u^5+u)}}, & x_2 &=\frac{\sqrt{u^4+1} z_4}{u},\\
x_1 &= -2 \ln (u), & x_3 &=\frac{\sqrt{u^4+1} z_3}{u}.
\end{align*}

\subsubsection{\JL\ models corresponding to $\{5;-1|2\}$}

Formula \eqref{E0spec} for the matrix \eqref{11to512} gives
$$ \hat E_0(t)=\left(
\begin{array}{cccc}
 0 & -t & 0 & 0 \\
 -t & 0 & 0 & 0 \\
 0 & 0 & \frac{1}{2} & \frac{1}{2} \\
 0 & 0 & -\frac{1}{2} & \frac{1}{2} \\
\end{array}
\right). $$ 
From the algebra $\{5;-1|2\}$ we get
$$ \hat \cf(t, x) =\left(
\begin{array}{cccc}
 0 & -e^{\frac{x_1}{2}} t & 0 & 0 \\
 -e^{\frac{x_1}{2}} t & 0 & 0 & 0 \\
 0 & 0 & \frac{e^{x_1}}{e^{2 x_1}+1} &
   \frac{1}{e^{2 x_1}+1} \\
 0 & 0 & -\frac{1}{e^{2 x_1}+1} &
   \frac{e^{x_1}}{e^{2 x_1}+1} \\
\end{array}
\right)$$
that together with  dilaton
$$
\hat \Phi(x)= -\half\ln \left(1+e^{2 x_1}\right)
$$
satisfy usual Supergravity Equations. The metric can be brought to Brinkmann form \eqref{ppwave12} by coordinate transformation that differs from the previous one only in sign of $x_1$.

\subsection{Plurality with one spectator - plurals to curved model on $\{1|1\}$}\label{11specIVO}

Even though it is easier to find initial \JL\ model with vanishing dilaton one can find model with non-vanishing one as well. After a bit tedious calculations we were able to find $E_0(t)$ as
\begin{equation}\label{E0ivo}
E_0(t) = \left(
\begin{array}{cccc}
 0 &  F'(t) & 0 & 0 \\
 F'(t) & F(t) & 0 & 0 \\
 0 & 0 & 1 & 0 \\
 0 & 0 & 0 & 1 \\
\end{array}
\right).
\end{equation}
Choosing $F(t)=t$ we get \bkg\ 
\begin{equation}
\cF(t,x) = \left(
\begin{array}{cccc}
 0 & e^{-\frac{x_1}{2}} & 0 &
   0 \\
 e^{-\frac{x_1}{2}} &
   e^{-x_1}t & 0 & 0 \\
 0 & 0 & e^{-x_1} & 0 \\
 0 & 0 & 0 & e^{-x_1} \\
\end{array}
\right).
\end{equation}
\sugra s are satisfied for rather exotic dilaton
\begin{equation}
\Phi(x)=-e^{e^{-\frac{x_1}{2}}}
   \text{Ei}\left(-e^{-\frac{x_1}{2}}\right
   )-\frac{x_1}{2}
\end{equation}
where the exponential integral function $\text{Ei}$ is defined as
$$\text{Ei}(x)=\int_{-\infty}^x \frac{e^t}{t} d t.$$
This model can serve for testing the formula \eqref{hatFAxdep} as 
$$
F_A=\left(0,e^{e^{-\frac{x_1}{2}}-\frac{x_1}{2}} \text{Ei}\left(-e^{-\frac{x_1}{2}}   \right)+1,0,0,0,0,0,0\right)
$$
depends on group \coor s.

Scalar curvature and torsion vanish. Metric in the Brinkmann form
\begin{equation}\label{ppwave51b}
ds^2 =-\frac{z_3^2+z_4^2 }{u^2 \ln (u)}du^2 + 2dudv + dz_3^2 + dz_4^2  
\end{equation}
can be obtained by 
\begin{align*}
t &= -\frac{2 u v \ln (u)+v^2+z_3^2}{4 \ln (u)}, & x_2 &=\frac{z_3}{\ln (u)},  \\
x_1 &=-2 \ln (\ln (u)), & x_3 &= \frac{v}{\ln (u)}.
\end{align*}

\subsubsection{ \JL\ models corresponding to $\{3;-2|1\}$ }

Formula \eqref{E0spec} for the matrix \eqref{11to321}
gives
$$ \hat E_0(t)=\left(
\begin{array}{cccc}
 0 & -2 & 0 & 0 \\
 -2 & 4t & 0 & 0 \\
 0 & 0 & \frac{5}{4} & \frac{3}{4} \\
 0 & 0 & \frac{3}{4} & \frac{5}{4} \\
\end{array}
\right). $$ 
From the algebra $\{3;-2|1\}$ we get
$$
\hat \cf(t, x) =\left(
\begin{array}{cccc}
 0 & -2 e^{x_1} & 0 & 0 \\
 -2 e^{x_1} & 4 e^{2 x_1} t & 0 &
   0 \\
 0 & 0 & e^{-2 x_1}+\frac{e^{2 x_1}}{4} &
   e^{-2 x_1}-\frac{e^{2 x_1}}{4} \\
 0 & 0 & e^{-2 x_1}-\frac{e^{2 x_1}}{4} &
   e^{-2 x_1}+\frac{e^{2 x_1}}{4} \\
\end{array}
\right) $$
that together with dilaton
$$
\Phi(x)= - e^{e^{x_1}} \text{Ei}\left(-e^{x_1}\right)$$
obtained from the formula \eqref{specdil393} satisfy Supergravity Equations. The curvature scalars as well as the first $\alpha'$ correction to the \sugra s vanish so that this \bkg\ is probably a pp-wave even though we were not able to find the transformation to the Brinkmann form. 

\subsubsection{\JL\ models corresponding to $\{5;-1|1\}$   }
Formula \eqref{E0spec} for the matrix \eqref{11to511} and the algebra $\{5;-1|1\}$ yield
$$ \hat \cf(t, x) =
\left(
\begin{array}{cccc}
 0 & -e^{\frac{x_1}{2}}& 0 & 0 \\
 -e^{\frac{x_1}{2}}  & e^{x_1}t &
   0 & 0 \\
 0 & 0 & e^{-x_1} & 0 \\
 0 & 0 & 0 & e^{-x_1} \\
\end{array}
\right) $$
that together with dilaton
$$ \Phi=- e^{\frac{x_1}{2}} \text{Ei}\left(-e^{\frac{x_1}{2}}\right) -\frac{x_1}{2}$$
satisfy Supergravity Equations. Scalar curvature of the metric is zero.

Metric in the Brinkmann form
\begin{equation}
ds^2 =\frac{\left(z_3^2+z_4^2\right) (\ln (u)+2)}{u^2 \ln ^2(u)}du^2 + 2dudv + dz_3^2 + dz_4^2  
\end{equation}
can be obtained by 
\begin{align*}
t &= \frac{-2 u v \ln (u)+z_3^2+z_4^2}{4 \ln (u)}, & x_2 &= z_4 \ln (u),  \\
x_1 &= 2 \ln (\ln (u)), & x_3 &= z_3 \ln (u).
\end{align*}

\subsubsection{\JL\ models corresponding to $\{1|2\}$}
For the matrix \eqref{11to12} and algebra $\{1|2\}$ we get

$$ \hat \cf(t, x) =\left(
\begin{array}{cccc}
 0 & e^{-\frac{x_1}{2}} & 0 & 0 \\
 e^{-\frac{x_1}{2}} & e^{-x_1} t &
   0 & 0 \\
 0 & 0 & \frac{e^{x_1}}{e^{2 x_1}+1} &
   \frac{1}{e^{2 x_1}+1} \\
 0 & 0 & -\frac{1}{e^{2 x_1}+1} &
   \frac{e^{x_1}}{e^{2 x_1}+1} \\
\end{array}
\right) $$
that together with dilaton
$$\Phi(x)= -
   e^{-\frac{x_1}{2}}
   \text{Ei}\left(-e^{-\frac{x_1}{2}}\right)
   +\frac{x_1}{2}$$
satisfy Supergravity Equations. Scalar curvature vanishes. Metric in the Brinkman form
\begin{equation}\label{ppwave12ivo}
ds^2 =\frac{\left(z_3^2+z_4^2\right) \left(\ln ^8(u)+2 \ln ^7(u)-10 \ln ^3(u)-1\right)}{u^2 \ln (u)
   \left(\ln ^4(u)+1\right)^2}du^2 + 2dudv + dz_3^2 + dz_4^2  
\end{equation}
can be obtained by transformation
$$
t = -\frac{2 u v \ln ^5(u)+2 u v \ln (u)-\left(z_3^2+z_4^2\right) \ln ^4(u)+z_3^2+z_4^2}{4 \left(\ln ^5(u)+\ln (u)\right)},$$
$$
x_1 = -2 \ln (\ln (u)), \quad x_2 = \frac{z_4 \sqrt{\ln ^4(u)+1}}{\ln (u)},\quad  x_3 = \frac{z_3 \sqrt{\ln ^4(u)+1}}{\ln (u)}.
$$

\subsubsection{\JL\ models corresponding to $\{5;-1|2\}$}

For the matrix \eqref{11to512} and algebra $\{5;-1|2\}$ we get
$$ \hat \cf(t, x) =\left(
\begin{array}{cccc}
 0 & -e^{\frac{x_1}{2}}  & 0 & 0 \\
 -e^{\frac{x_1}{2}} & e^{x_1} t &
   0 & 0 \\
 0 & 0 & \frac{e^{x_1}}{e^{2 x_1}+1} &
   \frac{1}{e^{2 x_1}+1} \\
 0 & 0 & -\frac{1}{e^{2 x_1}+1} &
   \frac{e^{x_1}}{e^{2 x_1}+1} \\
\end{array}
\right)$$
that together with  dilaton
$$\Phi(x)=-e^{\frac{x_1}{2}}
   \text{Ei}\left(-e^{\frac{x_1}{2}}\right)
  +\frac{x_1}{2}$$
satisfy Supergravity Equations. Scalar curvature vanishes. We can bring the metric to the Brinkman form \eqref{ppwave12ivo} by a coordinate transformation that differs from the previous one only in sign of $x_1$.

\subsection{Plurality with two spectators - plurals to sigma model on $[1|1]$}

To verify formula \eqref{specdil393} it is desirable to check it also for more than one spectator. In the following we show an example of \jltpy\ with two spectators $y_1,y_2$ and model  \crspto two-dimensional \JL\ bialgebra.

\JL\ bialgebra $[1|1]$ is formed by Abelian Lie algebras $\cg, \tcg$ and cocycle $\phi_0 = 2 Z_a T^a = T^{1}+T^{2}$. Choosing
\begin{equation}
E_0(y_1,y_2)= \left(
\begin{array}{cccc}
 0 & F(y_2) H(y_1) & 0 & 0 \\
 F(y_2) H(y_1) & 0 & 0 & 0 \\
 0 & 0 & 2 & 0 \\
 0 & 0 & 0 & 1 \\
\end{array}
\right)
\end{equation}
and
$$\varphi=\frac{1}{4} \left(x_1+x_2\right), \quad \hat d = -\half \ln \left(F(y_2) H(y_1) \right)$$ 
we get flat \JL\ model with vanishing dilaton $\Phi = 0$.

\subsubsection{\JL\ model corresponding to $[2;-1|1]$}

Isomorphism $ [1|1]\cong [2;-1|1]$  given by \eqref{c1121}  
transforms $E_0(y_1,y_2)$ to 
$$ \hat E_0(y_1,y_2)=\left(
\begin{array}{cccc}
 0 & F(y_2) H(y_1) & 0 & 0 \\
 F(y_2) H(y_1) & 0 & 0 & 0 \\
 0 & 0 & \frac{1}{3} & -\frac{2}{3} \\
 0 & 0 & \frac{2}{3} & \frac{2}{3} \\
\end{array}
\right). $$
\jltpy\ gives \bkg 
\begin{equation}\label{cf2spectators}
\hat \cf(x,y)=\left(
\begin{array}{cccc}
 0 & F(y_2) H(y_1) & 0 & 0 \\
 F(y_2) H(y_1) & 0 & 0 & 0 \\
 0 & 0 & \frac{e^{x_2}}{3} & \frac{1}{3} e^{x_2}
   \left(x_1-2\right) \\
 0 & 0 & \frac{1}{3} e^{x_2} \left(x_1+2\right) &
   \frac{1}{3} e^{x_2} \left(x_1^2+2\right) \\
\end{array}
\right)
\end{equation}
with nontrivial scalar curvature. Together with dilaton
$$\Phi(x,y)=-\half x_2$$
obtained from \eqref{specdil393} they satisfy \sugra s.

\begin{table}
\begin{center}
{\renewcommand{\arraystretch}{1.6}
\begin{tabular}{|c || c | c | c |}
\hline
Algebra & $K(u)$ & Torsion & Dilaton  \\
\hline \hline  
$ \{1|1\} $ & 0 & $ 0$ &0 \\
\hline
$ \{5;-1|1\} $ & $\frac{2 }{u^2} $ & $ 0$ &$-2\ln \left(u\right)$ \\
\hline
$ \{5;-1|2\} $ & $\frac{2 u^2 \left(u^4-5\right) }{\left(u^4+1\right)^2} $ & $\frac{4\,u}{1+u^4}(du\wedge dz_3\wedge dz_4)$ & $ -\frac{1}{2} \ln \left(u^4+1\right) $ \\
\hline
$ \{1|2\} $ & $ \frac{2 u^2 \left(u^4-5\right) }{\left(u^4+1\right)^2} $ & $-\frac{4\,u}{1+u^4}(du\wedge dz_3\wedge dz_4)$&$-\frac{1}{2} \ln \left(u^4+1\right)$ \\
\hline
\end{tabular}
}
\normalsize
\caption{Plane-parallel waves \eqref{pp-wave} with $K(u,\vec z) = K(u)( z_3^2+ z_4^2)$ obtained as four-dimensional \JL\ models in Sec. \ref{11spec}. 
\label{table5}}
\end{center}
\end{table}

\begin{table}
\begin{center}
{\renewcommand{\arraystretch}{1.6}
\begin{tabular}{|c || c | c | c |}
\hline
Algebra & $ K(u)$  & Torsion & Dilaton  \\
\hline \hline  
$ \{1|1\} $ & $-\frac{1}{u^2 \ln (u)}$& $ 0$ &$\ln (\ln (u))-u \text{Ei}(-\ln (u))$ \\
\hline
$ \{5;-1|1\} $ & $\frac{(\ln (u)+2) }{u^2 \ln ^2(u)} $ & $ 0$ &$-u\, \text{Ei}(-\ln (u))-\ln (\ln (u))$ \\
\hline
$ \{5;-1|2\} $ & $\frac{\left(\ln ^8(u)+2 \ln ^7(u)-10 \ln ^3(u)-1\right)}{u^2 \ln (u)
   \left(\ln ^4(u)+1\right)^2}$ & $\frac{4 \ln (u)}{u+u \ln ^4(u)}
(du\wedge dz_3\wedge dz_4)$ & $ \frac{1}{4} \ln \left(\frac{e^{-4 u \text{Ei}(-\ln
   (u))}}{\left(\frac{1}{\ln ^4(u)}+1\right)^2 \ln ^4(u)}\right) $ \\
\hline
$ \{1|2\} $ & $ \frac{\left(\ln ^8(u)+2 \ln ^7(u)-10 \ln ^3(u)-1\right)}{u^2 \ln (u) \left(\ln ^4(u)+1\right)^2} $ & $-\frac{4 \ln (u)}{u+u \ln ^4(u)}(du\wedge dz_3\wedge dz_4)$  & $\frac{1}{4} \ln \left(\frac{e^{-4 u \text{Ei}(-\ln
   (u))}}{\left(\frac{1}{\ln ^4(u)}+1\right)^2 \ln ^4(u)}\right)$ \\
\hline
\end{tabular}
}
\normalsize
\caption{Plane-parallel waves \eqref{pp-wave} with $K(u,\vec z) = K(u)( z_3^2+ z_4^2)$ obtained as four-dimensional \JL\ models in Sec. \ref{11specIVO}. 
\label{table6}}
\end{center}
\end{table}
\newpage

\section{Conclusions}

In this paper we have studied \jltpy\ as a solution-generating technique to Supergravity Equations. Focusing on Type 1 Leibniz algebras, i.e. those with ${f_b}^{ba} = 0$ and $X_0 = Z^a T_a  = 0$, we have found isomorphisms among four- and six-dimensional \JL\ bialgebras classified in \cite{rezaseph:class}. These isomorphisms allow us to distinguish Type 1 algebras into several equivalence classes, see Sec. \ref{sec:iso}, where \jltpy\ transformation can be applied.

For each equivalence class we have found an initial \JL\ model that satisfies \sugra s  \eqref{betaG}--\eqref{betaPhi} and used \JL\ T-plurality to transform it to other \JL\ models satisfying \sugra s. Examples of this procedure are given in Sections \ref{examples} and \ref{sec:spec}. 

It is usually assumed in the literature that initial flux $F_A$ is constant. For models with $F_A$ dependent on group \coor s we have used slightly modified formula \eqref{hatFAxdep} for their \tfn, and checked it for pluralities in Sections \ref{sec1|1}, \ref{11specIVO}. Working with non-constant $F_A$ was allowed by particular forms of $C$-matrices that did not bring dependency on dual coordinates in $\hat F_A$. Beside that we have fixed function $\hat d(y)$ in \eqref{dilmelsaka} for models with spectators as 
$$
\hat d(y)=-\frac{1}{4}\ln\frac{(\det E_0(y))^2}{\det E_{0S}(y)}, \quad E_{0S}=\half(E_0+E_0^T)
$$
to obtain formula for dilaton. All plural \bkg s obtained this way satisfy \sugra s.

Due to the particular choice of initial \JL\ models  all plural \bkg s 
 except \eqref{cf2spectators} have vanishing scalar curvature and simple Ricci tensor. For some of them we were able to bring them to the Brinkmann form, see Tables \ref{table4}, \ref{table5} and \ref{table6} to prove  that they  are plane-parallel waves. Besides that we have checked that the first $\alpha'$ correction to the \sugra s \cite{HT,borsatowulf,iranci} vanishes for all \bkg s with vanishing scalar curvature. That indicates that all \JL\ \bkg s with vanishing scalar curvature  derived in this paper are probably pp-waves \cite{ts}.

In some cases \JL\ T-plurality is a mere coordinate \tfn, see \bkg s in sections \ref{sec1|1}, \ref{412}, \ref{512}, in others it relates different pp-waves, see \eqref{ppwave4} and \eqref{ppwave34}. 

Let us note that the \bkg s presented above could not be simply obtained by \PL T-plurality for which $\Delta=\omega=0$ and the factor $\Exp{-2\omega}$ in \eqref{JLmtz} is trivial. In cases where one would get a flat model by \PL T-plurality,  atomic \jltpy\ produces conformally flat models.

\section{Appendix - Isomorphisms of Type 1 algebras}

Below we give an overview of particular isomorphisms relating different four- and six-dimensional \JL\ bialgebras via \eqref{Cplurality}. For notation and relation to classification carried out in \cite{rezaseph:class} see Section \ref{sec:iso}.

\subsection{Four-dimensional algebras}

\begin{equation*}
[1|1] \rightarrow [2;-1|1]\ : \ \
C= \left(
\begin{array}{cccc}
 0 & 0 & 0 & 1 \\
 -1 & 0 & 0 & 0 \\
 -1 & 1 & 0 & 0 \\
 0 & 0 & -1 & -1 \\
\end{array}
\right),
\end{equation*} 
\begin{equation*}
[2;\alpha|1] \rightarrow [2;-\frac{\alpha}{\alpha+1}|1]\ : \ \
C= \left(
\begin{array}{cccc}
 0 & 0 & 1 & 0 \\
 0 & -\frac{1}{\alpha +1} & 0 & 0 \\
 1 & 0 & 0 & 0 \\
 0 & 0 & 0 & -\alpha -1 \\
\end{array}
\right).
\end{equation*}

\subsection{Six-dimensional algebras}

\subsubsection{$C$-\mtr ces for pluralities \eqref{1|1}}

\begin{equation*}
 \{1\,|\,1\}
\rightarrow \{3;-2\,|\,1\}\ : \ \ C= 
\left(
\begin{array}{cccccc}
 -2 & 0 & 0 & 0 & 0 & 0 \\
 0 & 0 & \frac{1}{2} & 0 & 1 & 0 \\
 0 & 0 & -\frac{1}{2} & 0 & 1 & 0 \\
 0 & 0 & 0 & -\frac{1}{2} & 0 & 0 \\
 0 & \frac{1}{2} & 0 & 0 & 0 & 1 \\
 0 & \frac{1}{2} & 0 & 0 & 0 & -1 \\
\end{array}
\right),
\end{equation*}
\begin{equation*} 
\{1\,|\,1\}
\rightarrow\{5;-1\,|\,1\}\ : \ \ C= 
\left(
\begin{array}{cccccc}
 -1 & 0 & 0 & 0 & 0 & 0 \\
 0 & 0 & 0 & 0 & 1 & 0 \\
 0 & 0 & 0 & 0 & 0 & 1 \\
 0 & 0 & 0 & -1 & 0 & 0 \\
 0 & 1 & 0 & 0 & 0 & 0 \\
 0 & 0 & 1 & 0 & 0 & 0 \\
\end{array}
\right),
\end{equation*}
\begin{equation*}
\{1\,|\,1\}
\rightarrow \{1\,|\,2\}
 \ : \ \ C= 
\left(
\begin{array}{cccccc}
 1 & 0 & 0 & 0 & 0 & 0 \\
 0 & 1 & 0 & 0 & 0 & 0 \\
 0 & 0 & 1 & 0 & 0 & 0 \\
 0 & 0 & 0 & 1 & 0 & 0 \\
 0 & 0 & -1 & 0 & 1 & 0 \\
 0 & 1 & 0 & 0 & 0 & 1 \\
\end{array}
\right),
\end{equation*}
\begin{equation*}
\{1\,|\,1\}
\rightarrow \{5;-1\,|\,2\}\ : \ \ C= 
\left(
\begin{array}{cccccc}
 -1 & 0 & 0 & 0 & 0 & 0 \\
 0 & 0 & 0 & 0 & 1 & 0 \\
 0 & 0 & 0 & 0 & 0 & -1 \\
 0 & 0 & 0 & -1 & 0 & 0 \\
 0 & 1 & 0 & 0 & 0 & 1 \\
 0 & 0 & -1 & 0 & 1 & 0 \\
\end{array}
\right).
\end{equation*}

\subsubsection{$C$-\mtr ces for pluralities \eqref{2|1}}

\begin{equation*}
\{2\,|\,1\}
\rightarrow \{4;-1\,|\,1\}\ : \ \ C= 
\left(
\begin{array}{cccccc}
 0 & 0 & -1 & 0 & 0 & 0 \\
 0 & 0 & 0 & 1 & 0 & 0 \\
 0 & 0 & 0 & 0 & -1 & 0 \\
 0 & 0 & 0 & 0 & 0 & -1 \\
 1 & 0 & 0 & 0 & 0 & 0 \\
 0 & -1 & 0 & 0 & 0 & 0 \\
\end{array}
\right),
\end{equation*}
\begin{equation*}
\{2\,|\,1\}
\rightarrow \{2.i\,|\,1\}\ : \ \ C= 
\left(
\begin{array}{cccccc}
 0 & 0 & 1 & 0 & 0 & 0 \\
 1 & 0 & 0 & 0 & 0 & 0 \\
 0 & -1 & 0 & 0 & 0 & 0 \\
 0 & 0 & 0 & 0 & 0 & 1 \\
 0 & 1 & 0 & 1 & 0 & 0 \\
 1 & 0 & 0 & 0 & -1 & 0 \\
\end{array}
\right),
\end{equation*}
\begin{equation*}
\{2\,|\,1\}
\rightarrow \{2.ii\,|\,1\}\ : \ \ C= 
\left(
\begin{array}{cccccc}
 0 & 0 & 1 & 0 & 0 & 0 \\
 1 & 0 & 0 & 0 & 0 & 0 \\
 0 & 1 & 0 & 0 & 0 & 0 \\
 0 & 0 & 0 & 0 & 0 & 1 \\
 0 & -1 & 0 & 1 & 0 & 0 \\
 1 & 0 & 0 & 0 & 1 & 0 \\
\end{array}
\right),
\end{equation*}
\begin{equation*}
\{2\,|\,1\}
\rightarrow \{3:-2\,|\,2\}\ : \ \ C= 
\left(
\begin{array}{cccccc}
 0 & 0 & -2 & 0 & 0 & 0 \\
 \frac{1}{2} & 0 & 0 & 0 & -2 & 0 \\
 -\frac{1}{2} & 0 & 0 & 0 & -2 & 0 \\
 0 & 0 & 0 & 0 & 0 & -\frac{1}{2} \\
 0 & -\frac{1}{4} & 0 & 1 & 0 & 0 \\
 0 & -\frac{1}{4} & 0 & -1 & 0 & 0 \\
\end{array}
\right),
\end{equation*}
\begin{equation*}
\{2\,|\,1\}
\rightarrow \{4;-1\,|\,2\}\ : \ \ C= 
\left(
\begin{array}{cccccc}
 0 & 0 & -1 & 0 & 0 & 0 \\
 0 & 0 & 0 & 1 & 0 & 0 \\
 0 & 0 & 0 & 0 & -1 & 0 \\
 0 & 0 & 0 & 0 & 0 & -1 \\
 1 & 0 & 0 & 0 & 1 & 0 \\
 0 & -1 & 0 & 1 & 0 & 0 \\
\end{array}
\right),
\end{equation*}
\begin{equation*}
\{2\,|\,1\}
\rightarrow \{4iii;1\,|\,2\}\ : \ \ C= 
\left(
\begin{array}{cccccc}
 0 & 0 & 1 & 0 & 0 & 0 \\
 0 & 0 & 0 & 1 & 0 & 0 \\
 0 & 0 & 0 & 0 & -1 & 0 \\
 0 & 0 & 0 & 0 & 0 & 1 \\
 1 & 0 & 0 & 0 & -1 & 0 \\
 0 & -1 & 0 & -1 & 0 & 0 \\
\end{array}
\right).
\end{equation*}

\subsubsection{$C$-\mtr ces for pluralities \eqref{31|1}}

\begin{equation*}
\{3;-1\,|\,1\}
\rightarrow \{3;2\,|\,1\}\ : \ \ C=\left(
\begin{array}{cccccc}
 -\frac{1}{2} & 0 & 0 & 0 & 0 & 0 \\
 0 & -\frac{1}{2} & \frac{1}{2} & 0 &
   -\frac{1}{2} & -\frac{1}{2} \\
 0 & \frac{1}{2} & -\frac{1}{2} & 0 &
   -\frac{1}{2} & -\frac{1}{2} \\
 0 & 0 & 0 & -2 & 0 & 0 \\
 0 & -1 & 0 & 0 & 0 & 1 \\
 0 & -1 & 0 & 0 & 0 & -1 \\
\end{array}
\right),
\end{equation*}
\begin{equation*}
\{3;-1\,|\,1\}
\rightarrow \{6_0;1\,|\,1\}\ : \ \ C= 
\left(
\begin{array}{cccccc}
 0 & 0 & 0 & 0 & 1 & 0 \\
 0 & 0 & 0 & 0 & 0 & 1 \\
 -1 & 0 & 0 & 0 & 0 & 0 \\
 0 & 1 & 0 & 0 & 0 & 0 \\
 0 & 0 & 1 & 0 & 0 & 0 \\
 0 & 0 & 0 & -1 & 0 & 0 \\
\end{array}
\right),
\end{equation*}
\begin{equation*}
\{3;-1\,|\,1\}
\rightarrow \{6_0iii;1\,|\,2\}\ : \ \ C=   \left(
\begin{array}{cccccc}
 -1 & 0 & 0 & 0 & 0 & 0 \\
 0 & 0 & 0 & 0 & 0 & 1 \\
 0 & 0 & 0 & 0 & -1 & 0 \\
 0 & 0 & 0 & -1 & 0 & 0 \\
 0 & 0 & 1 & 0 & 1 & 0 \\
 0 & -1 & 0 & 0 & 0 & 1 \\
\end{array}
\right).
\end{equation*}

\subsubsection{$C$-\mtr ces for pluralities \eqref{3b|1}}

$$ b\neq -2,-1,2 $$
\begin{equation*}
\{3;b\,|\,1\}
\rightarrow \{3;\frac{-2b}{2+b}\,|\,1\}\ : \ \ C= 
\left(
\begin{array}{cccccc}
 -\frac{2}{b+2} & 0 & 0 & 0 & 0 & 0 \\
 0 & \frac{1}{4} & -\frac{1}{4} & 0 & 1 & 1 \\
 0 & -\frac{1}{4} & \frac{1}{4} & 0 & 1 & 1 \\
 0 & 0 & 0 & -\frac{b}{2}-1 & 0 & 0 \\
 0 & \frac{1}{4} & \frac{1}{4} & 0 & 1 & -1 \\
 0 & \frac{1}{4} & \frac{1}{4} & 0 & -1 & 1 \\
\end{array}
\right),
\end{equation*}
\[ \{3;b\,|\,1\}
\rightarrow \{3;b\,|\,2\}\ : \ \ C= 
\left(
\begin{array}{cccccc}
 1 & 0 & 0 & 0 & 0 & 0 \\
 0 & 1 & 0 & 0 & 0 & 0 \\
 0 & 0 & 1 & 0 & 0 & 0 \\
 0 & 0 & 0 & 1 & 0 & 0 \\
 0 & 0 & -\frac{1}{b+2} & 0 & 1 & 0 \\
 0 & \frac{1}{b+2} & 0 & 0 & 0 & 1 \\
\end{array}
\right),
  \]
  \[ \{3;b\,|\,1\}
\rightarrow \{3;\frac{-2b}{2+b}\,|\,2\}\ : \ \ C= 
\left(
\begin{array}{cccccc}
 -\frac{2}{b+2} & 0 & 0 & 0 & 0 & 0 \\
 0 & -\frac{1}{2} & \frac{1}{2} & 0 &
   -\frac{2}{b+2} & -\frac{2}{b+2} \\
 0 & \frac{1}{2} & -\frac{1}{2} & 0 &
   -\frac{2}{b+2} & -\frac{2}{b+2} \\
 0 & 0 & 0 & -\frac{b}{2}-1 & 0 & 0 \\
 0 & \frac{1}{4} (-b-2) & 0 & 0 & 0 & 1 \\
 0 & \frac{1}{4} (-b-2) & 0 & 0 & 0 & -1 \\
\end{array}
\right),
  \]
  \[ \{3;b\,|\,1\}
\rightarrow \{6_{b+1};-b\,|\,1\}\ : \ \ C= 
\left(
\begin{array}{cccccc}
 -1 & 0 & 0 & 0 & 0 & 0 \\
 0 & 0 & 0 & 0 & 1 & 0 \\
 0 & 0 & 0 & 0 & 0 & 1 \\
 0 & 0 & 0 & -1 & 0 & 0 \\
 0 & 1 & 0 & 0 & 0 & 0 \\
 0 & 0 & 1 & 0 & 0 & 0 \\
\end{array}
\right),
  \]
  \[ \{3;b\,|\,1\}
\rightarrow \{6_{\frac{2-b}{2+b}};\frac{2b}{2+b}\,|\,1\}\ : \ \ C= 
\left(
\begin{array}{cccccc}
 \frac{2}{b+2} & 0 & 0 & 0 & 0 & 0 \\
 0 & \frac{1}{4} & \frac{1}{4} & 0 & -1
   & 1 \\
 0 & \frac{1}{4} & \frac{1}{4} & 0 & 1 &
   -1 \\
 0 & 0 & 0 & \frac{b}{2}+1 & 0 & 0 \\
 0 & -\frac{1}{4} & \frac{1}{4} & 0 & 1
   & 1 \\
 0 & \frac{1}{4} & -\frac{1}{4} & 0 & 1
   & 1 \\
\end{array}
\right),
  \]  \[ \{3;b\,|\,1\}
\rightarrow \{6_{b+1};-b\,|\,2\}\ : \ \ C= 
\left(
\begin{array}{cccccc}
 -1 & 0 & 0 & 0 & 0 & 0 \\
 0 & 0 & 0 & 0 & 1 & 0 \\
 0 & 0 & 0 & 0 & 0 & 1 \\
 0 & 0 & 0 & -1 & 0 & 0 \\
 0 & 1 & 0 & 0 & 0 & -\frac{1}{b+2} \\
 0 & 0 & 1 & 0 & \frac{1}{b+2} & 0 \\
\end{array}
\right),
  \]
  \[ \{3;b\,|\,1\}
\rightarrow \{6_{\frac{2-b}{2+b}};\frac{2b}{2+b}\,|\,2\}\ : \ \ C= 
\left(
\begin{array}{cccccc}
 \frac{2}{b+2} & 0 & 0 & 0 & 0 & 0 \\
 0 & \frac{1}{b+2} & \frac{1}{b+2} & 0 & 1 & -1
   \\
 0 & \frac{1}{b+2} & \frac{1}{b+2} & 0 & -1 & 1
   \\
 0 & 0 & 0 & \frac{b}{2}+1 & 0 & 0 \\
 0 & 0 & -\frac{1}{2} & 0 & \frac{b}{2}+1 & 0 \\
 0 & 0 & \frac{1}{2} & 0 & \frac{b}{2}+1 & 0 \\
\end{array}
\right).
  \]

\subsubsection{$C$-\mtr ces for pluralities \eqref{4b|1}}

$$ b\neq -2, -1 $$
\[ \{4;b\,|\,1\}
\rightarrow \{4;\frac{-b}{1+b}\,|\,1\}\ : \ \ C= 
\left(
\begin{array}{cccccc}
 -b-1 & 0 & 0 & 0 & 0 & 0 \\
 0 & 0 & 0 & 0 & 0 & 1 \\
 0 & 0 & 0 & 0 & b+1 & 0 \\
 0 & 0 & 0 & -\frac{1}{b+1} & 0 & 0 \\
 0 & 0 & 1 & 0 & 0 & 0 \\
 0 & \frac{1}{b+1} & 0 & 0 & 0 & 0 \\
\end{array}
\right),
  \]
 
\[ \{4;b\,|\,1\}
\rightarrow  \{4;b\,|\,2\}\ : \ \ C= 
\left(
\begin{array}{cccccc}
 1 & 0 & 0 & 0 & 0 & 0 \\
 0 & 1 & 0 & 0 & 0 & 0 \\
 0 & 0 & 1 & 0 & 0 & 0 \\
 0 & 0 & 0 & 1 & 0 & 0 \\
 0 & 0 & -\frac{1}{b+2} & 0 & 1 & 0 \\
 0 & \frac{1}{b+2} & 0 & 0 & 0 & 1 \\
\end{array}
\right),
  \]
 
\[ \{4;b\,|\,1\}
\rightarrow  \{4;\frac{-b}{1+b}\,|\,2\}\ : \ \ C= 
\left(
\begin{array}{cccccc}
 -b-1 & 0 & 0 & 0 & 0 & 0 \\
 0 & 0 & 0 & 0 & 0 & b+2 \\
 0 & 0 & 0 & 0 & b^2+3 b+2 & 0 \\
 0 & 0 & 0 & -\frac{1}{b+1} & 0 & 0 \\
 0 & 0 & \frac{1}{b+2} & 0 & -b-1 & 0 \\
 0 & \frac{1}{b^2+3 b+2} & 0 & 0 & 0 & 1 \\
\end{array}
\right).
  \]

\subsubsection{$C$-\mtr ces for pluralities \eqref{5b|1}}

$$ b\neq -2,-1 $$
\[ \{5;b\,|\,1\}
\rightarrow \{5;\frac{-b}{1+b}\,|\,1\}\ : \ \ C= 
\left(
\begin{array}{cccccc}
 1 & 0 & 0 & 0 & 0 & 0 \\
 0 & 1 & 0 & 0 & 0 & 0 \\
 0 & 0 & 1 & 0 & 0 & 0 \\
 0 & 0 & 0 & 1 & 0 & 0 \\
 0 & 0 & -\frac{1}{b+2} & 0 & 1 & 0 \\
 0 & \frac{1}{b+2} & 0 & 0 & 0 & 1 \\
\end{array}
\right),
  \]
 
\[ \{5;b\,|\,1\}
\rightarrow \{5;b\,|\,2\}\ : \ \ C= 
\left(
\begin{array}{cccccc}
 1 & 0 & 0 & 0 & 0 & 0 \\
 0 & 1 & 0 & 0 & 0 & 0 \\
 0 & 0 & 1 & 0 & 0 & 0 \\
 0 & 0 & 0 & 1 & 0 & 0 \\
 0 & 0 & -\frac{1}{b+2} & 0 & 1 & 0 \\
 0 & \frac{1}{b+2} & 0 & 0 & 0 & 1 \\
\end{array}
\right),
\]

\[ \{5;b\,|\,1\}
\rightarrow \{5;\frac{-b}{1+b}\,|\,2\}\ : \ \ C= 
\left(
\begin{array}{cccccc}
 -\frac{1}{b+1} & 0 & 0 & 0 & 0 & 0 \\
 0 & 0 & 0 & 0 & 0 & \frac{b+2}{b+1} \\
 0 & 0 & 0 & 0 & -\frac{b+2}{b+1} & 0 \\
 0 & 0 & 0 & -b-1 & 0 & 0 \\
 0 & 0 & \frac{b+1}{b+2} & 0 & 1 & 0 \\
 0 & -\frac{b+1}{b+2} & 0 & 0 & 0 & 1 \\
\end{array}
\right).  
\] 

\subsubsection{$C$-\mtr ces for pluralities \eqref{60b1}}

\[ \{6_0;b\,|\,1\}
\rightarrow \{6_0.iii;b\,|\,2\}\ : \ \ C= 
\left(
\begin{array}{cccccc}
 0 & 0 & 1 & 0 & 0 & 0 \\
 0 & 1 & 0 & 0 & 0 & 0 \\
 -1 & 0 & 0 & 0 & 0 & 0 \\
 0 & 0 & 0 & 0 & 0 & 1 \\
 \frac{1}{b} & 0 & 0 & 0 & 1 & 0 \\
 0 & \frac{1}{b} & 0 & -1 & 0 & 0 \\
\end{array}
\right),
\]
\[ \{7_0;b\,|\,1\}
\rightarrow \{7_0.i;b\,|\,2\}\ : \ \ C= 
\left(
\begin{array}{cccccc}
 0 & 0 & 1 & 0 & 0 & 0 \\
 0 & 1 & 0 & 0 & 0 & 0 \\
 1 & 0 & 0 & 0 & 0 & 0 \\
 0 & 0 & 0 & 0 & 0 & 1 \\
 -\frac{1}{b} & 0 & 0 & 0 & 1 & 0 \\
 0 & \frac{1}{b} & 0 & 1 & 0 & 0 \\
\end{array}
\right),
\]
\[ \{7_0;b\,|\,1\}
\rightarrow \{7_0.ii;b\,|\,2\}\ : \ \ C= 
\left(
\begin{array}{cccccc}
 0 & 0 & 1 & 0 & 0 & 0 \\
 0 & 1 & 0 & 0 & 0 & 0 \\
 -1 & 0 & 0 & 0 & 0 & 0 \\
 0 & 0 & 0 & 0 & 0 & 1 \\
 \frac{1}{b} & 0 & 0 & 0 & 1 & 0 \\
 0 & \frac{1}{b} & 0 & -1 & 0 & 0 \\
\end{array}
\right).
\]

\end{document}